\documentclass[acmsmall,screen]{acmart}
\AtBeginDocument{%
  }

\setcopyright{acmlicensed}
\copyrightyear{2026}
\acmYear{2026}
\acmDOI{}
\acmISBN{}

\acmConference[TECS]{ACM Trans. Embedd. Comput. Syst.}{Sept. 17--09, 2026}{New York, NY, USA}
\makeatletter
\gdef\acmConference@shortname{\@gobble}%
\gdef\acmConference@venue{}%
\makeatother
\acmBooktitle{ACM. Trans. Embedd. Comput. Syst.\ Sept. 17--09, 2026}

\usepackage[most]{tcolorbox}
\usepackage{subcaption}
\usepackage{booktabs}
\usepackage{tabularx}
\usepackage{wrapfig}
\usepackage{amsmath}
\usepackage[ruled,vlined,linesnumbered]{algorithm2e}
\usepackage{ragged2e}   
\newcolumntype{L}{>{\RaggedRight\arraybackslash}X}
\usepackage{multirow}
\usepackage[table]{xcolor} 
\usepackage{array}
\usepackage{pifont}
\usepackage{indentfirst}
\usepackage{array, booktabs, multirow, cellspace, makecell, tabularx}
\usepackage{graphicx} 
\usepackage[table]{xcolor} 
\usepackage{flushend} 

\usepackage{booktabs,tabularx,array,multirow}
\usepackage{colortbl}
\usepackage{graphicx}
\usepackage{caption}
\usepackage{tcolorbox}
\tcbuselibrary{listings,skins}

\definecolor{kwblue}{RGB}{0,102,153}
\definecolor{cmtmaroon}{RGB}{128,0,0}
\lstdefinestyle{akiraC}{
  language=C,
  basicstyle=\ttfamily\scriptsize,
  keywordstyle=\color{kwblue},
  commentstyle=\color{cmtmaroon},
  stringstyle=\ttfamily,          
  numbers=left,
  numberstyle=\tiny\color{black},
  numbersep=4pt,
  frame=tb,
  backgroundcolor=,               
  breaklines=true,
  showstringspaces=false,
  columns=fullflexible,
  morekeywords={fabric_bridge_set_isolation,fabric_gate,
                wrpkru,rdpkru,pkru_set_rw,pkru_set_none,
                vmept_set_func,vmept_wait_for_rpc}
} 

\newcommand{\cmark}{\checkmark}
\newcommand{\xmark}{\ding{55}}

\newcolumntype{L}[1]{>{\raggedright\arraybackslash}p{#1}}
\newcolumntype{C}[1]{>{\centering\arraybackslash}p{#1}}

\newcolumntype{Y}{>{\raggedright\arraybackslash}X} 
\newcolumntype{C}[1]{>{\centering\arraybackslash}m{#1}} 
\newcolumntype{L}[1]{>{\raggedright\arraybackslash}m{#1}} 
\tcbset{
  insightbox/.style={
    colback=white,
    colframe=black,
    boxrule=0.7pt,
    arc=2pt,
    left=4pt,
    right=4pt,
    top=3pt,
    bottom=3pt,
    boxsep=0pt,
    fonttitle=\bfseries,
    before skip=6pt,
    after skip=6pt
  }
}

\makeatletter
\AtBeginDocument{%
  \fancypagestyle{standardpagestyle}{%
    \fancyhf{}%
    \renewcommand{\headrulewidth}{\z@}%
    \renewcommand{\footrulewidth}{\z@}%
    \fancyhead[LE]{\@headfootfont\@acmArticlePage}%
    \fancyhead[RO]{\@headfootfont\@acmArticlePage}%
    \fancyhead[RE]{\@headfootfont\@shortauthors}%
    \fancyhead[LO]{\@headfootfont\shorttitle}%
  }%
  \fancypagestyle{firstpagestyle}{%
    \fancyhf{}%
    \renewcommand{\headrulewidth}{\z@}%
    \renewcommand{\footrulewidth}{\z@}%
    \fancyhead[LE]{\@acmBadgeL}%
    \fancyhead[LO]{\@acmBadgeL}%
    \fancyhead[RO]{\@acmBadgeR}%
    \fancyhead[RE]{\@acmBadgeR}%
  }%
  \pagestyle{standardpagestyle}%
}
\makeatother 

\begin{document}

\title{Akira: Towards Operating System Architecture for Agile Specialization}

\author{Renshuang Jiang}
\email{renshuang717@163.com}
\orcid{1234-5678-9012}
\authornotemark[1]
\affiliation{%
  \institution{National University of Defense Technology}
  \city{ChangSha}
  \state{HuNan}
  \country{China}
}

\author{Pan Dong}
\email{pandong@nudt.edu.cn}
\orcid{1234-5678-9012}
\affiliation{%
  \institution{National University of Defense Technology}
  \city{ChangSha}
  \state{HuNan}
  \country{China}
}

\author{Yichong Wang}
\email{yichongwang@nudt.edu.cn}
\orcid{1234-5678-9012}
\affiliation{%
  \institution{National University of Defense Technology}
  \city{ChangSha}
  \state{HuNan}
  \country{China}
}

\author{Lei Li}
\email{3433324076@qq.com }
\orcid{1234-5678-9012}
\affiliation{%
  \institution{National University of Defense Technology}
  \city{ChangSha}
  \state{HuNan}
  \country{China}
}

\author{Yuchen Dong}
\email{YuchenWorks@yeah.net}
\orcid{1234-5678-9012}
\affiliation{%
  \institution{Durham University}
  \country{United Kingdom}
}

\author{Xiaoxiang Fang}
\email{xiaoxiang200@163.com}
\orcid{1234-5678-9012}
\affiliation{%
  \institution{National University of Defense Technology}
  \city{ChangSha}
  \state{HuNan}
  \country{China}
}
\author{Qirui Yu}
\email{yuqirui@nudt.edu.cn}
\orcid{1234-5678-9012}
\affiliation{%
  \institution{National University of Defense Technology}
  \city{ChangSha}
  \state{HuNan}
  \country{China}
}

\author{Jianfeng Zhang}
\email{jfzhang@nudt.edu.cn}
\orcid{1234-5678-9012}
\affiliation{%
  \institution{National University of Defense Technology}
  \city{ChangSha}
  \state{HuNan}
  \country{China}
}

\author{Zhe Jiang}
\email{zhejiang.uk@gmail.com}
\authornotemark[2]
\orcid{1234-5678-9012}
\affiliation{%
  \institution{Southeast University}
  \city{ChangSha}
  \state{HuNan}
  \country{China}
}

\newcommand{\zhenote}[1]{\textbf{\textcolor{blue}{zhe: #1}}}
\newcommand{\shuangnote}[2]{\textbf{\textcolor{magenta}{shuang: #1}}}
 
\renewcommand{\shortauthors}{Jiang et al.}

\begin{abstract}
Operating Systems (OS) specialization tailors abstractions and privileges to specific use cases, enabling configurable tradeoffs between performance and isolation. 
Modern OS designs adopt modular architectures (e.g., microkernels and LibOSs) that decompose OSs into components for customization. 
However, increasing application diversity makes low-cost and non-invasive agile specialization challenging. 
First, modular OS designs decompose the system into components and reduce coupling, but lack abstractions separating component implementations from OS configuration.
Developers must still reason about cross-component dependencies and inspect module internals. 
Second, diverse workloads require different isolation strengths, yet existing OSs largely fix isolation models at design time, making isolation adaptation changes costly and invasive. 

To address these, we present Akira, an OS architecture that enables efficiently adapt OS composition and isolation. 
Akira introduces the Abstraction and Encapsulation Unit (AEU), which separates component interfaces and isolation requirements from implementations, hiding modular OS internals and enabling non-invasive reconfiguration. 
Based on the abstraction, Akira introduces the Control Logic Unit (CLU) and the Bridge-based Isolation Unit (BIU) to separately manage component communication and isolation, enabling their independent adaptation without system-wide modifications.  
Specifically, CLU separates component management from component functionality through explicit routing and control policies, enabling communication updates without modifying component code. 
BIU groups components with similar isolation needs into shared domains and assigns isolation mechanisms accordingly, allowing each domain to match isolation to its performance and security needs.  
As a demonstration, we prototype Akira using FreeRTOS and Unikraft as components and privilege/MPU/MPK/EPT as isolation on representative hardware. 
Results show that Akira enables rapid and low-cost OS reconfiguration with only a 1.10$\times$ slowdown over native FreeRTOS and reduces reconfiguration effort from 1.38 KLOC of intrusive changes to 0.28 KLOC of incremental updates. On Unikraft-based Redis and Nginx, Akira configurations expand the performance–isolation space by up to 10.4$\times$ and 7.8$\times$. 
We publicly release Akira’s source code, at https://github.com/JiangRS17/Akira. 
\end{abstract}

\ccsdesc[500]{Computer systems organization~Architectures}

\keywords{Operating System, Agile Specialization, Abstraction, Configurable Isolation }


\authorsaddresses{}
\maketitle

\section{Introduction}


General-purpose Operating Systems (OSs) offer extensive functionality and heavyweight mechanisms that can compromise performance and security when deployed in specialized environments~\cite{zhang2021demikernel,kuenzer2021unikraft}.
OS specialization~\cite{jung2026wayfinder, lankes2016hermitcore}, i.e., tailoring the OS to suit a specific application or use case, can yield orders-of-magnitude speedups or more robust security guarantees by customizing or removing abstractions and privileges. 
For instance, by bypassing the traditional network stack and providing direct access to packet I/O, a streamlined packet I/O framework achieves over 20× higher throughput than a standard Linux stack \cite{rizzo2012netmap}.  
Likewise, library OS designs enable application-specific Virtual Machines (VMs) that include only required OS functionalities, reducing unnecessary system components and the attack surface~\cite{kuenzer2021unikraft}. 
As deployment scenarios and workloads diversify, adapting an existing specialized OS to new requirements often requires extensive redesign and redevelopment. 
For example, ClickOS  is specialized for high-throughput packet processing. However, retargeting it to scenarios involving strongly isolated, multitenant services typically requires substantial redesign~\cite{martins2014clickos}. 
This creates a need for agile OS specialization, in which systems can be rapidly reconfigured with minimal redevelopment effort to meet disparate security and performance requirements.  

\begin{figure}[htpb]
\centerline{\includegraphics[width=\textwidth]{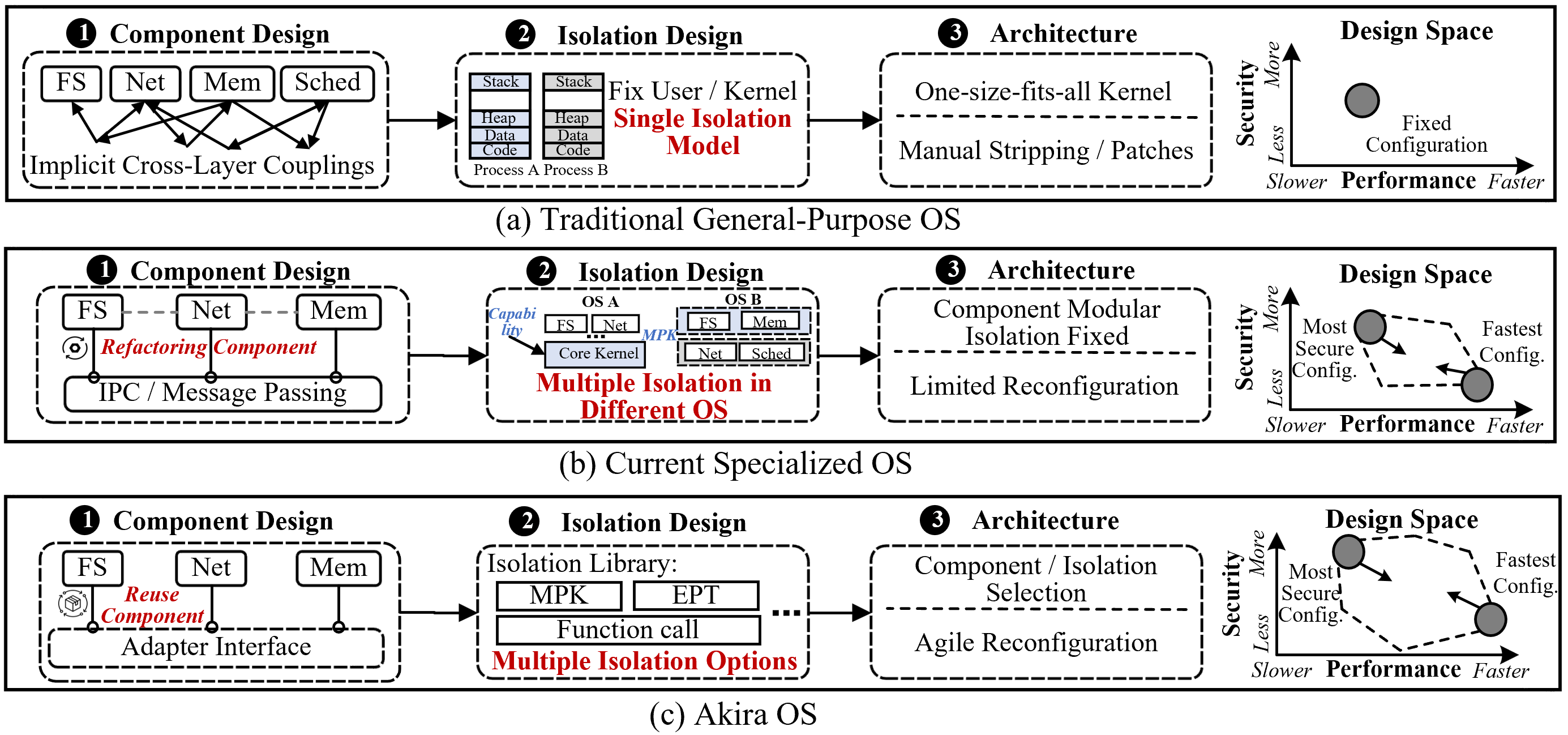}}
\vspace{-1.0em}
\caption{Comparison of OS specialization approaches.
(a) Traditional OS architectures rely on tightly coupled components and fixed isolation models, limiting customization to a single design point. (b) Existing specialized OSs improve modularity through component decomposition and refactoring, but isolation choices remain tied to specific architectures. (c) Akira reuses existing components, supports multiple isolation mechanisms within a single OS, and significantly expands the specialization design space. 
}
\label{intro}
\end{figure}

\noindent 
\textbf{Existing Work.} 
OS functionality is jointly realized through interactions among system components and enforcement of protection boundaries~\cite{boos2020theseus, narayanan2020redleaf, lefeuvre2022flexos, sartakov2021cubicleos}.  
Consequently, agile OS specialization necessarily involves coordinated changes in these two aspects, which require deep expertise in OS internals, including system components and isolation mechanisms, to balance trade-offs in performance and security. 
In Fig.~\ref{intro} (a), the larger codebases and tight couplings in traditional OS architectures mean that modifications to one subsystem often break assumptions in another, leading to system-wide changes~\cite{jacobsen2015lightweight}.  
Moreover, traditional OS architectures also have complex internal dependencies that make them difficult to understand and maintain, requiring developers to be familiar with low-level implementation details.  
For example, introducing full preemptive scheduling into Linux (to support real-time scenarios) required coordinated changes across numerous dependent subsystems, including scheduling, synchronization, and interrupt handling~\cite{reghenzani2019real,Github-RTpatch}. 
Beyond component coupling, these enforce fixed isolation boundaries~\cite{lefeuvre2022flexos, chen2024microkernel}.  
Strong isolation mechanisms, such as user/kernel separation, provide security at the cost of performance, whereas lightweight isolation, such as software fault isolation (SFI),  improves speed but reduces safety.  
Therefore, traditional OSs hard-code one point in this trade-off space\footnote{Here, the trade-off space refers to the choices among different points in the performance-security trade-off space, where improving one aspect may introduce costs in the other.} at design time, making post-deployment adjustments extremely difficult~\cite{lefeuvre2022flexos}. 
In Fig.~\ref{intro} (b), modular designs, such as microkernels and library OSs~\cite{klein2009sel4, kuenzer2021unikraft,madhavapeddy2013unikernels}, decompose OS functionality into independent components, expanding the design space of OS in different scenarios. 
However, although modularity improves component separation and reduces coupling, it cannot provide abstractions for hiding internal implementation details. 
Similarly, existing isolation provides diverse protection boundaries~\cite{watson2015cheri,hedayati2019hodor,feng2021scalable,gu2020harmonizing}, but remains unable to configure appropriate isolation for different requirements of the workload.

\noindent 
\textbf{Challenges.} 
Despite the expansion of design space with specialized OSs tailored to specific scenarios (Fig.~\ref{intro} (b)), rapidly adapting an OS to meet increasing and heterogeneous demands remains challenging. 
From the components’ perspective, modularity focuses on decomposing OS functionality into independent units and reducing component coupling. 
However, modularity mainly addresses the internal organization of components rather than the evolution of relationships. 
As systems evolve, new requirements often require changes to component interfaces, composition structures, and dependency relationships, which cannot be addressed by adding or removing modules. 
Developers must still reason about and coordinate these cross-component interactions manually, leading to invasive modifications when the system organization changes. 
For example, in seL4, IPC operation relies on strictly defined message formats. 
When interfaces evolve, such as introducing new fields, all communicating components must coordinate corresponding updates to maintain compatibility. 
From the isolation perspective, existing OS architectures make isolation choices at design time, coupling protection granularity with system organization.  
The static isolation and protection model in most OSs cannot satisfy the diverse isolation requirements of different use cases.  
Once an OS is built around a certain isolation model (e.g., user/kernel separation, process-based, or a single address space), changing the isolation model requires invasive modifications or a complete redesign.  
For example, introducing fine-grained isolation (isolating each component or library individually such as MPK~\cite{corbet2015mpk}, MPU~\cite{armv8mmpu}) into a coarse-grained protection (lumping subsystems behind the same boundaries such as user/kernel)  requires coordinated changes to memory management, privilege control, and communication. 
All of these limit an OS's agility in supporting rapid customization or reconfiguration.  
Although recent OSs such as FlexOS~\cite{lefeuvre2022flexos} provide more flexible isolation, their flexibility is achieved within a specific highly modular component organization (e.g., Unikraft-style library~OS). 
Applying this flexible isolation mechanism to platforms with different OSs still requires significant modifications (discussed in Sec.~\ref{motivation}).


\noindent \textbf{Contributions.}  
Here, we analyze the bottlenecks of current OS specialization, identify component abstraction and isolation modularity as key challenges, and extract insights to guide the design of a new OS architecture, i.e., Akira.   
In Fig.~\ref{intro}~(c), Akira introduces a stable abstraction layer for components that encapsulates interactions and hides implementation details. This abstraction enables components to be reused and evolved without modifying their internal implementations. 
To support diverse isolation requirements, Akira further provides multiple isolation options through an isolation library, so isolation domains can be configured with different protection mechanisms according to application needs. Isolation strategies can thus evolve together with component composition without redesigning the underlying components. 
At the architecture level, Akira introduces a controller that coordinates component and isolation selection, enabling agile reconfiguration of both. 
Overall, Akira provides an OS architecture for agile specialization that supports flexible component composition and isolation reconfiguration with minimal modifications to existing system components. 
As a demonstration, we prototype Akira with FreeRTOS and Unikraft and deploy it on a representative hardware platform. 
Experimental results show that Akira can be rapidly reconfigured to accommodate diverse workloads and requirements, incurring very light performance and communication overhead.

\section{Background and Motivation}
\label{section:motivation}  
With increasing diversity in workloads and deployment scenarios, OS specialization requires a flexible, low-cost, and efficient approach to agile adaptation~\cite{jung2026wayfinder}. 
In this section, we review existing approaches on OS specialization and analyze their limitations in supporting agile customization, from which we extract key architectural insights that motivate the design of a new OS architecture.   

\subsection{Mainstream OS Architectures and Features} 
Existing systems explore OS specialization through different architectural approaches. 
These can be broadly categorized into two principal implementation pathways, as summarized in Tab.~\ref{tab:os_comparison}. 

\begin{table}[htpb]
\centering
\caption{Comparison of existing OS designs relevant to agile specialization.}
\vspace{-0.5em}
\label{tab:os_comparison}
\resizebox{\textwidth}{!}{%
\begin{tabular}{@{}c|c|cc|cc|cc@{}}
\toprule
 &  & \multicolumn{2}{c}{Microkernel-based} & \multicolumn{2}{c}{Library OS} & \multicolumn{2}{c}{Hybrid Systems} \\
\cmidrule(r){3-4} \cmidrule(lr){5-6} \cmidrule(l){7-8}
\multirow{-2}{*}{\begin{tabular}[c]{@{}c@{}}OS \\ Specialization\end{tabular}}
 & \multirow{-2}{*}{Property}
& seL4~\cite{klein2009sel4}
 & \begin{tabular}[c]{@{}c@{}}Barrelfish~\cite{baumann2009multikernel}\\LegoOS~\cite{shan2018legoos}\end{tabular}
 & Exokernel~\cite{engler1995exokernel}
 & Unikraft~\cite{kuenzer2021unikraft}
 & FlexOS~\cite{lefeuvre2022flexos}
 & \begin{tabular}[c]{@{}c@{}}UnderBridge~\cite{gu2020harmonizing}\\HongMeng~\cite{chen2024microkernel}\end{tabular} \\
\midrule

 & Model
 & \begin{tabular}[c]{@{}c@{}}capability \\ based\end{tabular}
 & \begin{tabular}[c]{@{}c@{}}distributed / \\ multi-core isolation\end{tabular}
 & \begin{tabular}[c]{@{}c@{}}minimal kernel, \\ library-level\end{tabular}
 & \begin{tabular}[c]{@{}c@{}}single address \\ space\end{tabular}
 & \begin{tabular}[c]{@{}c@{}}hardware \\ assisted\end{tabular}
 & \begin{tabular}[c]{@{}c@{}}microkernel \end{tabular} \\
\cmidrule(l){2-8}

\multirow{-3}{*}{\begin{tabular}[c]{@{}c@{}}Isolation \\ Capabilities\end{tabular}}
 & Configurable
 & Fixed & Fixed & App-defined & Limited & Tunable & Partial \\
\midrule

 & Composition Model
 & \begin{tabular}[c]{@{}c@{}}fixed kernel \\ interfaces\end{tabular}
 & \begin{tabular}[c]{@{}c@{}}fixed kernel \\ interfaces\end{tabular}
 & \begin{tabular}[c]{@{}c@{}}fine-grained \\ library \end{tabular}
 & \begin{tabular}[c]{@{}c@{}}fine-grained \\ component\end{tabular}
 & \begin{tabular}[c]{@{}c@{}}Unikraft-style \\ decomposition\end{tabular}
 & \begin{tabular}[c]{@{}c@{}}Hardware \\ separation\end{tabular} \\
\cmidrule(l){2-8}

\multirow{-3}{*}{\begin{tabular}[c]{@{}c@{}}Composability \\ \& \\ Extensibility\end{tabular}}
 & Adaptation Cost
 & High & High & Low & Low & Medium--High & Medium \\
\midrule

 & Portability
 & \xmark & \xmark & \xmark & \xmark & \xmark & Partial \\
\cmidrule(l){2-8}

\multirow{-2}{*}{\begin{tabular}[c]{@{}c@{}}Agile \\ Specialization\end{tabular}}
 & Isolation Decoupled
 & \xmark & \xmark & Partial & \xmark & \cmark & Partial \\
\bottomrule
\end{tabular}%
}
\end{table}

Microkernel-based systems provide structured modular boundaries by separating kernel functionality from system services, with systems such as seL4~\cite{klein2009sel4}, Barrelfish~\cite{baumann2009multikernel}, and LegoOS~\cite{shan2018legoos} extending this model with capability-based resource management and distributed isolation. 
However, these isolation boundaries are typically defined together with component organization and kernel interfaces, and adapting them requires substantial restructuring of components and execution paths (detailed in Sec.~\ref{motivation}), making post-deployment adaptation costly and invasive~\cite{kilpatrick2003privman}, which limits agile OS specialization. 
In contrast, component-oriented or library OS designs (e.g., Exokernel~\cite{engler1995exokernel}, Unikraft~\cite{kuenzer2021unikraft}) construct application-specific OS instances through flexible component selection and composition, allowing system functionality to be tailored to application requirements. 
However, components often execute within shared protection domains without strong isolation enforcement, and maintaining interfaces and dependencies across components requires substantial engineering effort. 
For example, Unikraft adopts a highly efficient single-address-space design but does not provide built-in support for strong isolation between components. 
As a result, introducing additional isolation mechanisms often requires substantial system modifications, increasing adaptation cost.

A natural idea is to combine both pathways to achieve agile OS specialization. 
For example, hybrid systems such as FlexOS~\cite{lefeuvre2022flexos}, UnderBridge~\cite{gu2020harmonizing}, and HongMeng~\cite{chen2024microkernel}  achieve it by pairing reusable components with hardware-assisted isolation.  
However, their flexibility is typically realized within a fixed host-OS organization (e.g., Unikraft-style decomposition in FlexOS), so applying the same isolation approach to platforms with different component structures still incurs substantial adaptation cost (detailed in Sec.~\ref{motivation}; qualitative comparison in Sec.~\ref{sec:related}).  
Next, we further analyze the fundamental challenges and derive the key insights for achieving agile OS specialization.

\subsection{Motivation}
\label{motivation} 
\subsubsection{\textbf{Lack of Abstraction across Component Modularity.}} 
\begin{figure}[htpb]
\begin{minipage}[t]{0.47\linewidth}
\vspace{10pt}
\begin{lstlisting}[
  caption={Splitting \texttt{net\_process} into parse/filter/forward with a new descriptor changes the call ABI and ripples to other components.},
  label={lst:iface-cascade},
  captionpos=b,
  basicstyle=\ttfamily\footnotesize,
  keywordstyle=\color{purple!70!black},
  commentstyle=\color{teal},
  columns=fullflexible,
  keepspaces=true,
  aboveskip=2pt,
  belowskip=2pt,
  xleftmargin=2pt,
  xrightmargin=2pt,
]
long net_process(struct sk_buff *skb) {
    return ip_input(skb);  // one ABI / callee
}
long net_process(struct pkt_desc *d) {
    parse(d); filter(d);   // new call chain
    return forward(d, d->port); 
}
\end{lstlisting}
\end{minipage}%
\hfill
\begin{minipage}[t]{0.5\linewidth}
\vspace{0pt}
\centering
\includegraphics[width=0.8\linewidth]{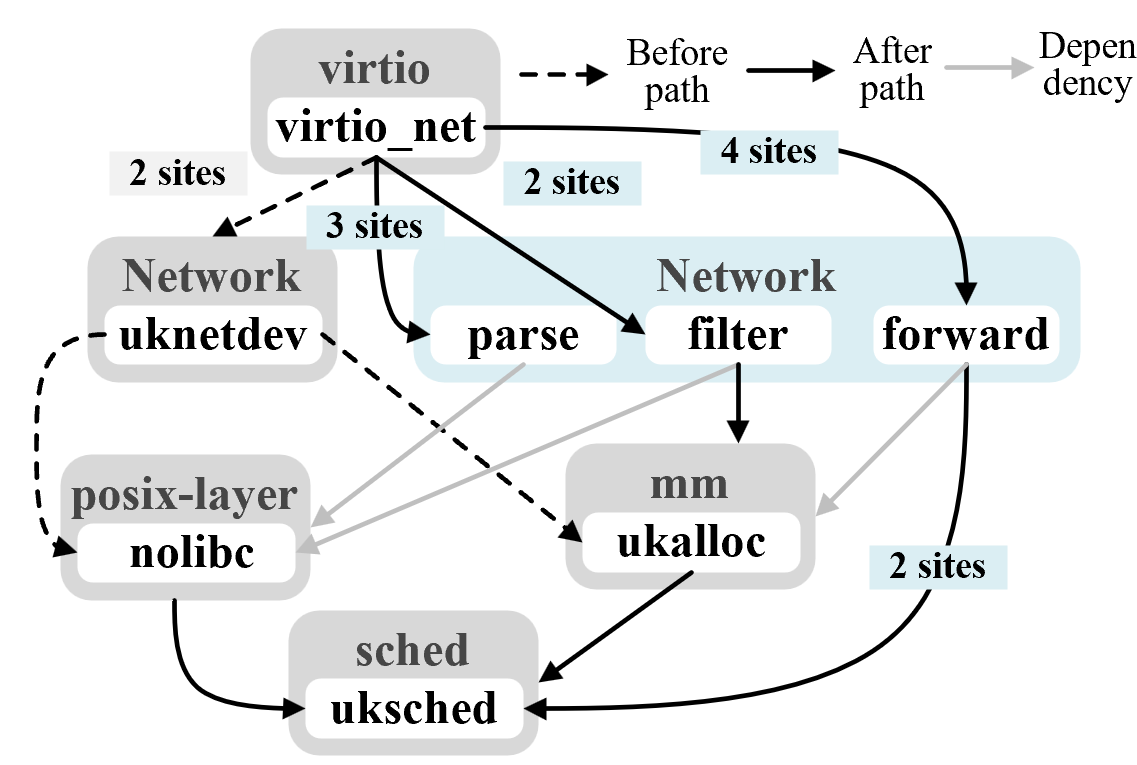}%
\vspace{-0.5em}
\captionof{figure}{Change Dependence.
( $n$ functions to update).}
\vspace{-0.5em}
\label{fig:deps}
\end{minipage}
\end{figure}

Component modularity helps decompose OS functionality into independent units and reduce coupling among components, providing a foundation for system composition and maintenance. 
However, the component organization in existing OSs is not a fixed configuration, but constantly evolves with changing functional requirements and hardware platforms.   
In particular, changes to external interfaces, calling conventions, and dependency relationships of components are common. 
As shown in Listing~\ref{lst:iface-cascade}, network packet processing evolves from a monolithic protocol into separate parsing, filtering, and forwarding modules~\cite{bosshart2014p4}, requiring changes to descriptor fields and invocation parameters. 
When interfaces change, existing bindings and compositions become invalid, prompting updates across all dependent callers (Fig.~\ref{fig:deps}). 
\textbf{Therefore, modularity alone cannot abstract away such evolving implementation complexity.}
Adding, replacing, or reconfiguring components still requires developers to understand implementation details and perform extensive manual modifications. 

To address these, a straightforward approach is to introduce abstraction layers.  
These layers encapsulate different interaction methods, calling conventions, and data representations into a unified interface before dispatching requests to target components. 
For example, CORBA~\cite{vinoski1997corba} provides interface definitions and request brokers to hide communication details between components. 
FlexOS~\cite{lefeuvre2022flexos} introduces gate mechanisms (e.g., \texttt{FlexOS\_gate}) to mediate cross-component interactions. 
However, existing abstractions often require developers to encode low-level organization decisions directly. CORBA uses IDL to describe interfaces and requires developers to coordinate invocations; FlexOS relies on annotations to specify compartment placement, permitted memory accesses, and APIs. Akira still requires developer input, but at a higher level: developers identify exposed interfaces and provide component-level isolation and backend constraints. Akira then utilizes these constraints to derive domains, select backends, and generate communication bridges, reducing manual coordination of the final deployment topology.

\begin{tcolorbox}[insightbox]
\small
\textbf{Key Insight \#1:} 
An abstraction layer is required to treat component details as black boxes, reducing implementation dependencies to support agile OS specialization. 
\end{tcolorbox}

\subsubsection{\textbf{Control Path Mixing in Modular OS Designs}} 
\label{obs2}
\begin{figure}[htbp]
\centering
\begin{minipage}{0.47\textwidth}
    \centering
    \includegraphics[width=\textwidth]{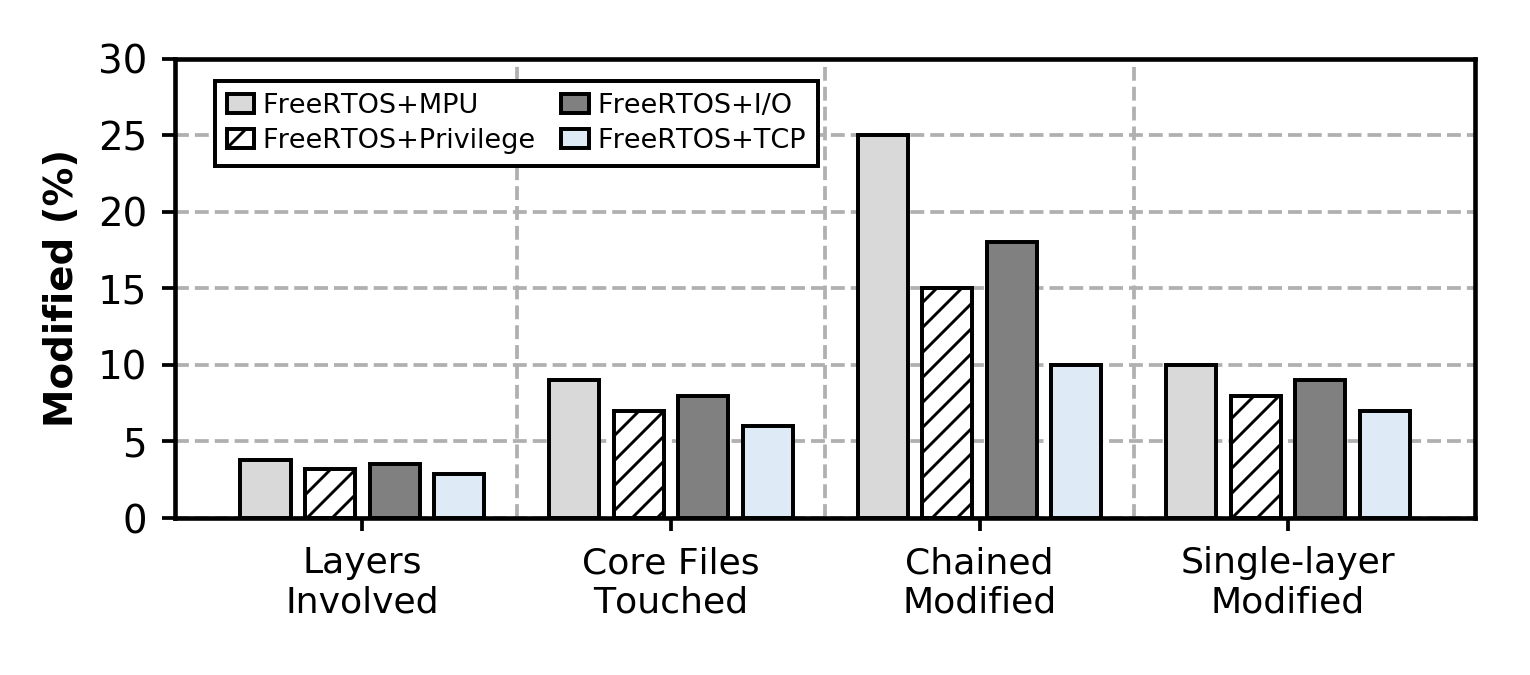}
    \vspace{-2.7em}
    \caption{Code-modification cost of adding isolation and I/O features in FreeRTOS.}
    \label{Akira_FreeRTOS}
\end{minipage}
\hfill
\begin{minipage}{0.5\textwidth}
    \centering
    \includegraphics[width=\textwidth]{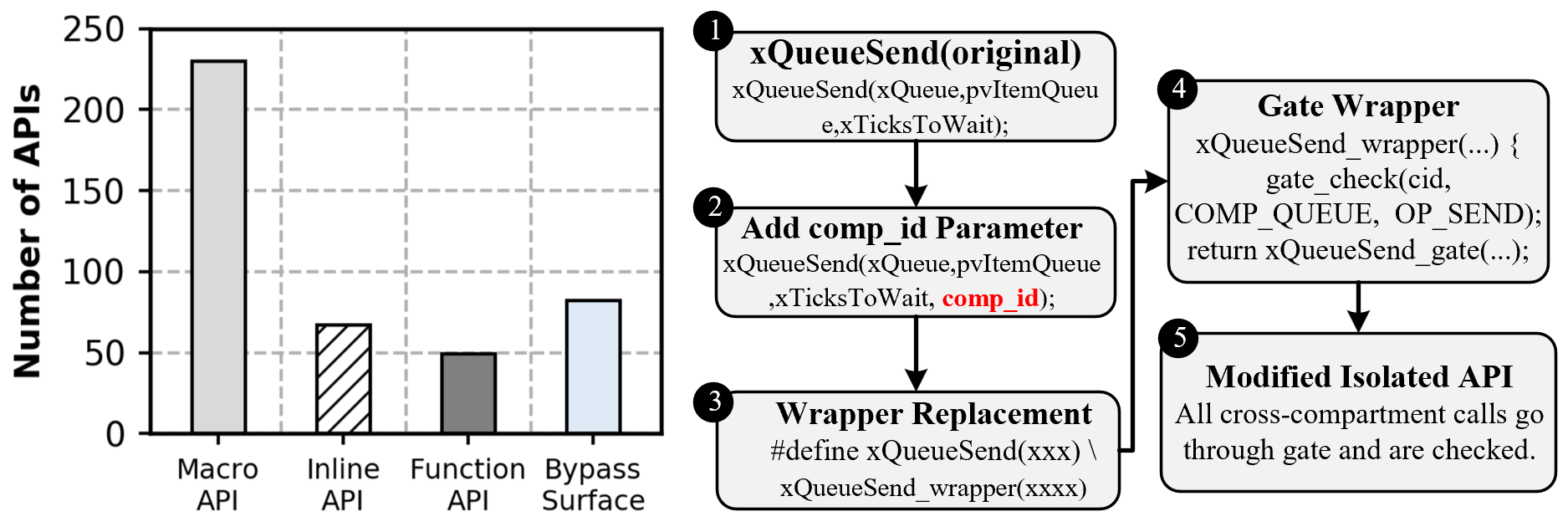}
    \vspace{-2.0em}
    \caption{FreeRTOS API statistics and a replacement workflow example.}
    \label{fig:akira_freertos_2}
\end{minipage}
\end{figure}

As discussed above, when components evolve internally, an abstraction layer is required to hide implementation details and prevent other components from being affected. 
However, even when component interfaces remain unchanged, existing modular designs still struggle to support low-cost adaptation when the system needs to change how components are isolated, communicate, or access resources.  
This is because policies governing isolation configuration, communication decisions, and resource management (hereafter referred to as control logic) are often mixed with component functionality and execution paths, requiring modifications across multiple components when system behaviors change. 
To quantify this, we analyze the effort of integrating new isolation mechanisms or optimizing I/O paths in FreeRTOS (Fig.~\ref{Akira_FreeRTOS}). 
This process typically spans 2.7–4.2 functional layers, requires modifications to 8.3 core files on average, and triggers cascading changes in 15\%–22\% of related interfaces; only $\leq$ 12\% of modifications can be confined within a single layer. 
These results indicate that existing isolation mechanisms are tightly coupled with control  structures, hindering low-cost adaptation. 

Furthermore, recent OS designs have attempted to separate control logic from component implementations by introducing configurable mechanisms.  
For example, FlexOS~\cite{lefeuvre2022flexos} and HongMeng~\cite{chen2024microkernel} provide configurable isolation mechanisms. 
However, these mechanisms remain coupled with the underlying OS architecture (e.g., FlexOS relies on Unikraft's decoupled component structure to realize hardware-enforced compartments).  
To quantify this, we migrate FlexOS-style isolation mechanisms to FreeRTOS and measure effort. 
In Fig.~\ref{fig:akira_freertos_2}, 
we classify FreeRTOS APIs according to where isolation logic needs to be integrated: API-level interfaces, compiler-expanded interfaces, kernel execution paths, and hardware-dependent paths, then quantify their proportions. 
Compiler-expanded interfaces are relatively easier to modify because they are visible at the source level and can be rewritten with macro definitions.  
However, even these interfaces require non-local changes, as modifications propagate to dependent execution paths and shared kernel operations. 
For example, enabling isolation for \texttt{xQueueSend} requires modifying 1543 lines across 9 files, including the macro interface, isolation dispatch, and queue operations. 
Other APIs directly interact with internal kernel paths or hardware mechanisms, making isolation integration require deeper kernel modifications.  

Overall, existing modular OS designs provide functional decomposition, but system behaviors are still governed by component implementations and execution paths, requiring invasive modifications rather than configuration changes for OS specialization.

\begin{tcolorbox}[insightbox]
\small
\textbf{Key Insight \#2:} 
A dedicated control module is required to separate system configuration and management logic from functional execution, enabling agile specialization without system-wide modifications. 
\end{tcolorbox}

\subsubsection{\textbf{Mismatch Between Fixed Isolation and Workload Diversity.}} 
Existing isolation mechanisms provide different protection boundaries, while introducing distinct performance and security trade-offs.
These can be classified based on protection boundaries (Tab.~\ref{Tab:sio_compare}), e.g., software fault isolation (SFI), domain-based isolation and privilege-based isolation.
These isolation mechanisms provide different protection capabilities while introducing distinct privilege contexts, memory access rules, and resource management constraints. 

\begin{table}[ht]
\centering
\caption{Isolation mechanisms classified by protection boundary. }
\vspace{-0.5em}
\label{Tab:sio_compare}
\setlength{\tabcolsep}{2.5pt}
\renewcommand{\arraystretch}{1.15}
\footnotesize
\begin{tabular}{@{}llcccc@{}}
\toprule
\makecell[c]{Protection\\boundary} &
Mechanism &
Enforcement &
\makecell[c]{Memory access} &
\makecell[c]{Cross-boundary\\switch} &
\makecell[c]{Overhead} \\
\midrule
Software fault &
MPX~\cite{intel2015mpx}, ASAN~\cite{serebryany2012addresssanitizer} &
Instr.\ checks &
Checked L/S in shared AS &
Inline checks &
High \\
\midrule
\multirow{3}{*}{Domain-based} &
MPK~\cite{corbet2015mpk} &
Protection keys &
Key-tagged page rights &
\texttt{wrpkru} &
Low \\
&
MPU~\cite{armv8mmpu} &
Region attributes &
Per-region R/W/X &
MPU reprogram &
Low \\
&
EPT~\cite{intel2015mpx} &
Nested paging &
GPA$\to$HPA mappings &
VMFUNC/VM-exit &
High \\
\midrule
Privilege-based &
User/Kernel &
CPU privilege levels &
User vs.\ kernel map &
syscall/sysret &
Med \\
\bottomrule
\end{tabular}
\end{table}

However, a general-purpose OS typically selects a single isolation mechanism at design time, assuming that it can satisfy diverse workload requirements. 
Therefore, the OS typically provides stronger protection mechanisms (e.g., privilege switch), but this imposes unnecessary overhead on lightweight computing tasks that require only domain-based protection. 
To evaluate the impact of such isolation choices, 
we characterize representative Embench~\cite{embench2020} workloads according to their protection requirements based on expert knowledge \footnote{The expert knowledge is derived from prior embedded-system design practices and the authors' experience in designing and evaluating FreeRTOS-based embedded systems.}  and established system design principles~\cite{saltzer1975protection} (Tab.~\ref{tab:isolation_recommendatio}), and measure their performance overhead under different isolation configurations in FreeRTOS (Fig.~\ref{observation1}). 
For \texttt{crc32}, \texttt{statemate}, and \texttt{ud} workloads, which mainly involve computation and data processing, MPU-based domain isolation is sufficient to prevent unintended memory accesses between components.  
Stronger isolation (e.g., privilege switch) increases overhead by 8\%, while providing limited additional protection benefits. 
For \texttt{huffbench}, \texttt{matmult-int} and \texttt{md5sum} workloads, which require strict control over execution contexts, privilege-based isolation is more suitable to prevent unauthorized execution. 
Insufficient isolation (e.g., None or MPU) may pose security risks to these workloads~\cite{feng2021scalable,amar2023cheriot}. 
These observations indicate that it is necessary to trade off performance and security across different workloads. 
The key limitation is not the lack of isolation, but that isolation is not treated as a composable property configurable to each workload.

\vspace{1.0em}
\begin{minipage}[!htb]{0.4\textwidth}
\captionof{table}{\centering Workload Security Requirements.}
    \centering
    \footnotesize
    \renewcommand{\arraystretch}{0.9}
    \vspace{-3mm} 
    \resizebox{\linewidth}{!}{%
    \begin{tabular}{lccc}
    \toprule
    Workload & Basic & Memory & HW Isolation \\
    \midrule
    crc32 & \ding{56} & \ding{52} & \ding{56} \\
    huffbench & \ding{56} & \ding{56} & \ding{52} \\
    matmult-int & \ding{56} & \ding{56} & \ding{52} \\
    md5sum & \ding{56} & \ding{56} & \ding{52} \\
    statemate & \ding{56} & \ding{52} & \ding{56} \\
    ud & \ding{56} & \ding{52} & \ding{56} \\
    \bottomrule
    \end{tabular}
    }
    \label{tab:isolation_recommendatio}
\end{minipage}
  \hfill
\begin{minipage}[!htb]{0.5\textwidth} 
    \centering
    \makebox[0.7\textwidth][c]{          
        \includegraphics[width=\textwidth]{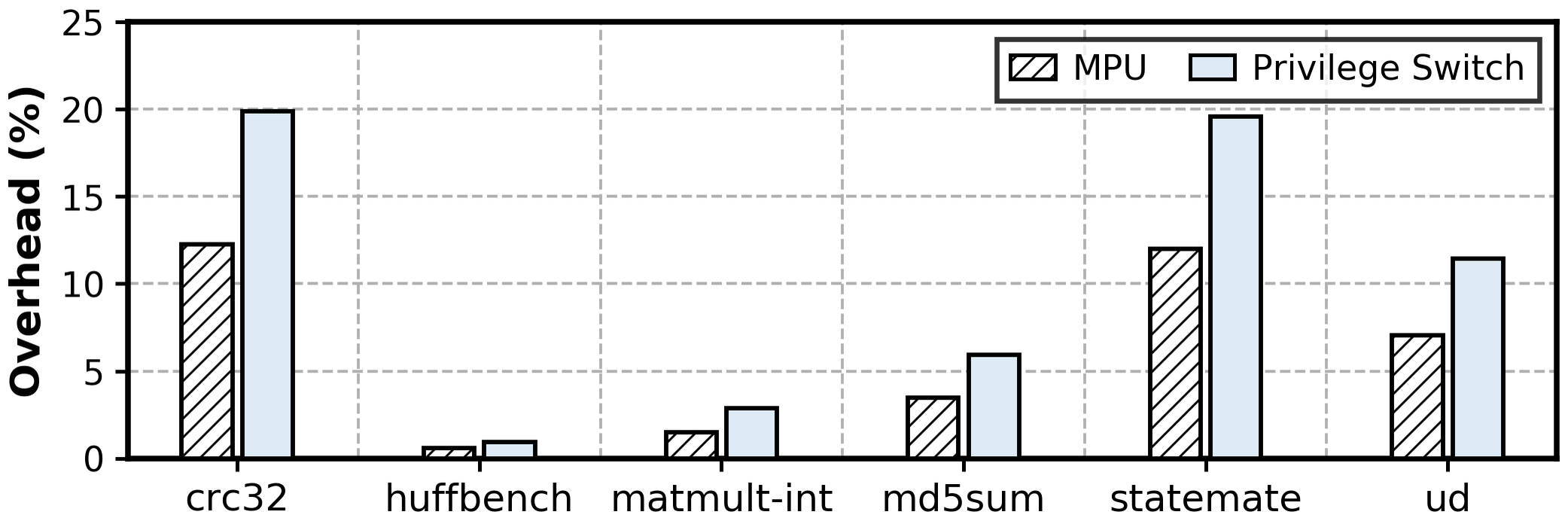} 
    }
    \vspace{-0.5em}
    \captionof{figure}{\centering Different Workloads Performance.}
    \vspace{-6pt}
    \label{observation1} 
\end{minipage}
\vspace{1em}

\begin{tcolorbox}[insightbox]
\small
\textbf{Key Insight \#3:} 
Fixed isolation granularity leads to mismatched protection and performance trade-offs across diverse workloads.
Isolation should therefore be treated as a composable architectural property, supporting multiple protection boundaries to accommodate different requirements. 
\end{tcolorbox}

\section{Akira Architecture Overview}
\begin{figure}[htpb]
\centerline{\includegraphics[width=\textwidth]{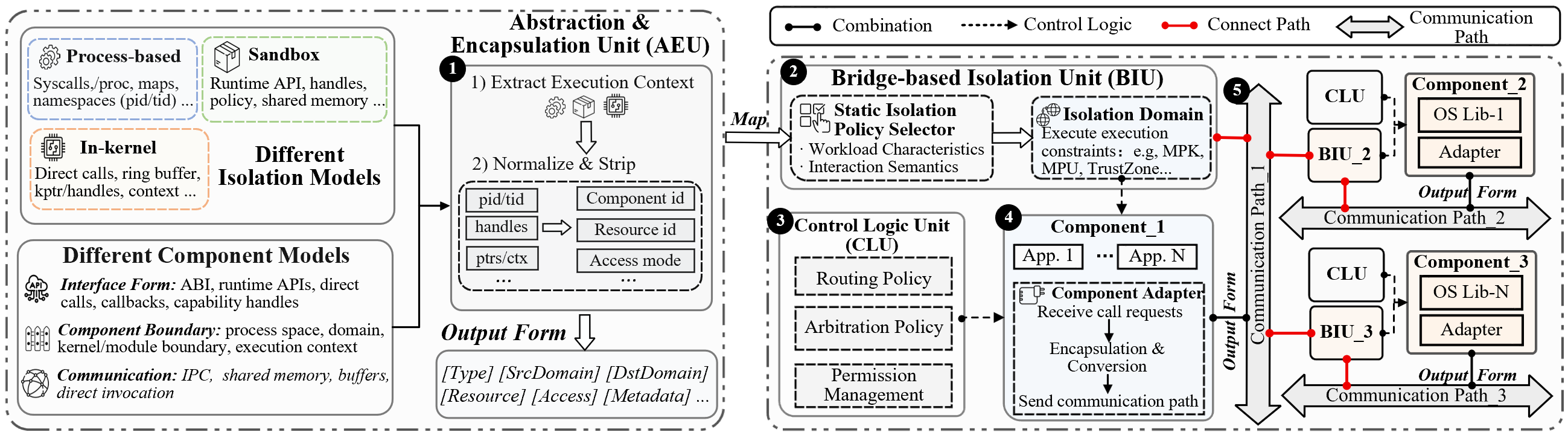}}
\caption{Akira Architecture Overview: AEU~\ding{182} extracts and normalizes execution and interaction semantics without modifying component implementations or isolation mechanisms; BIU~\ding{183} supports diverse isolation requirements by selecting and managing appropriate isolation domains for different workload characteristics; CLU~\ding{184} decouples interaction control from system execution paths and constructs routing. }
\vspace{-0.5em}
\label{akira_arch}
\end{figure}

In this paper, our goal is to support scenario-specific OS reconfiguration of components and isolation strategies, enabling low-cost and efficient system evolution and optimization. 
Based on these key insights, we propose \textit{Akira}, an OS architecture that establishes a stable abstraction for components and separates component functionality from system-level coordination and isolation management. 
First, Akira introduces an abstraction and encapsulation unit (AEU) to manage individual components by abstracting component interactions and isolation-related information into a unified representation while providing reusable abstraction libraries that hide component-specific details and enable component evolution without implementation modifications. 
Then, Akira introduces the control logic unit (CLU) to decouple communication and control logic from component functionality without requiring system-wide modifications. 
Besides, Akira designs the Bridge-based isolation unit (BIU) to group components with similar isolation requirements into the same domains and configure different isolation mechanisms, supporting diverse performance and security requirements. 
Therefore, Akira uses AEU to manage components and CLU/BIU to manage interactions between components, enabling agile OS reconfiguration without redesigning components or introducing new isolation mechanisms. 
Fig.~\ref{akira_arch} presents Akira's overview. 



To hide component-specific details and reduce implementation dependencies (Insight~\#1), Akira introduces an Abstraction and Encapsulation Unit (AEU) (Fig.~\ref{akira_arch} \ding{182}), which provides a common abstraction layer between components and underlying mechanisms. 
Component integration faces two fundamentally different challenges: adapting to the diversity of isolation mechanisms, where different approaches provide distinct protection boundaries and execution models, and the continuous evolution of component organizations, where interfaces, invocation patterns, and composition relationships may change over time. 
Therefore, Akira designs components and isolation abstractions separately in AEU. 
For isolation mechanisms, their configuration depends on protection requirements, boundaries, and execution conditions rather than enforcement details. Therefore, the AEU encapsulates information describing these aspects, including required protection levels, isolation boundaries, and contexts, while leaving concrete enforcement operations to the isolation mechanisms.  
For component organizations, their integration depends on how components interact. Therefore, the AEU extracts related information, including interface forms, invocation mechanisms, and resource access assumptions. 
This abstraction hides component- and isolation-specific details behind a common representation, enabling different mechanisms to be integrated, replaced, and evolved independently without propagating changes to component implementations or execution paths, while introducing only limited overhead during system construction (detailed in Sec.~\ref{evaluation}).

To avoid system-wide execution-path modifications (Insight~\#2), Akira designs a Control Logic Unit (CLU) as an independent interaction control abstraction that separates interaction decisions from kernel functional execution paths (Fig.~\ref{akira_arch} \ding{184}).  
Based on the descriptors provided by the AEU, the CLU manages component interactions through a communication fabric (Fig.~\ref{akira_arch} \ding{186}). 
The communication fabric represents interactions as explicit message exchanges rather than direct caller–callee invocations, while the CLU dynamically determines routing and dispatch decisions based on the provided descriptors. 
To realize these interaction control decisions, the CLU consists of a component registry table, a routing module, and a dispatch module, which together manage interaction control throughout the system lifetime. 
During component registration, the CLU extracts interaction metadata (e.g., interfaces and execution attributes) and builds corresponding routing entries in the registry table. 
During runtime, the CLU receives interaction requests, the routing module determines the target domain, and the dispatch module forwards requests along the path toward the BIU (responsible for subsequent permission checking and isolation enforcement).  
When component interfaces change, only the corresponding registry entries need to be updated, without modifying execution paths or control logic across the system.  
Besides, by moving interaction control outside component execution paths, CLU avoids embedding communication logic into individual components, reducing unnecessary processing overhead and improving system performance.

To support diverse isolation requirements and avoid fixed-granularity protection trade-offs (Insight~\#3), Akira introduces a Bridge-based Isolation Unit (BIU), which provides composable isolation through an isolation policy selector and a set of isolation domains (Fig.~\ref{akira_arch} \ding{183}). 
The selector determines suitable isolation mechanisms according to workload characteristics and interaction requirements, while the interaction descriptors generated by the AEU are used to configure corresponding isolation domains. 
Unlike systems that enforce a fixed isolation boundary for all components, BIU allows each component or interaction to be associated with an appropriate isolation domain based on its protection requirements. 
Based on the selected policies, isolation domains enforce protection and mediation for components participating in cross-domain interactions through the communication fabric. 
Each domain encapsulates its underlying isolation mechanism, allowing different protection boundaries to coexist within the same system. 
This design avoids forcing all components to adopt the same isolation granularity, enabling flexible protection–performance trade-offs according to workload requirements while reducing unnecessary isolation overhead.

Overall, Akira builds on existing modular system components by introducing a unified control and isolation architecture, including the AEU, CLU, and BIU mechanisms, thereby enabling isolation and component behaviors to evolve without system-wide modifications and providing a low-cost and flexible foundation for agile OS specialization. 
We provide a case study using an existing system to demonstrate this capability in Akira (Sec.~\ref{case_stduy}).

\section{Akira Design}
\label{implementation} 
Based on the proposed architecture, we implemented Akira as a prototype OS for agile OS specialization. 
It demonstrates the feasibility of the proposed design principles, but it is not the only way to build an agile OS. 
Akira is designed to realize three core capabilities: component abstraction for reusable functional representations (Sec.~\ref{aeu}), separation of component functionality from system-level communication and control logic through dedicated control units (Sec.~\ref{clu}), and flexible component composition into isolation domains for multi-level isolation (Sec.~\ref{biu}).

\subsection{Abstraction and Encapsulation Unit (AEU)} 
\label{aeu} 
From Insight \#1, adapting different isolation and component mechanisms to new workloads or requirements is costly and system-wide. 
Akira tackles this challenge by wrapping each component and isolation mechanism in an Abstraction and Encapsulation Unit (AEU), while leaving their internal implementations unchanged.

\begin{figure}[htpb]
\centerline{\includegraphics[width=0.7\textwidth]{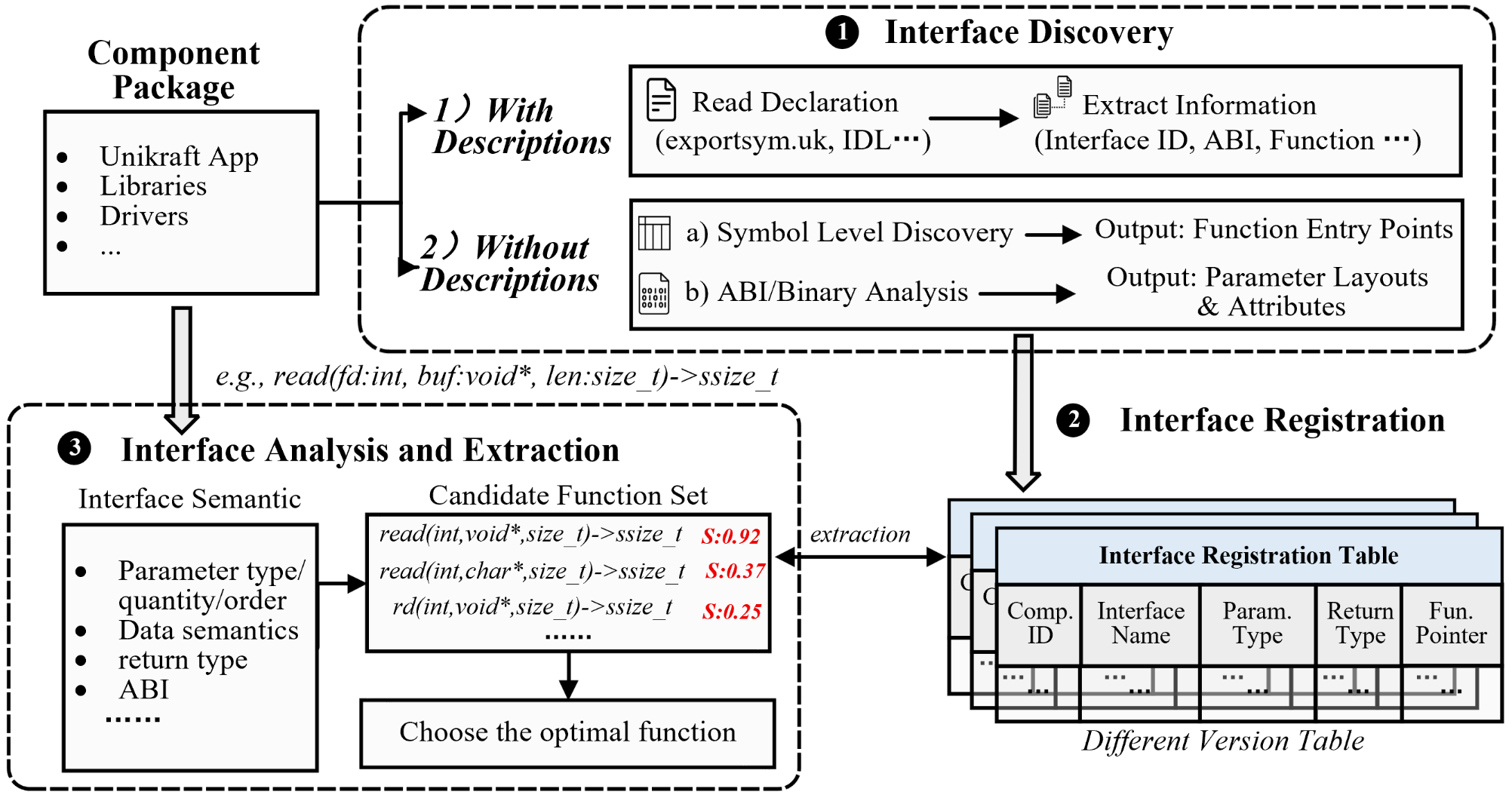}}
\vspace{-0.5em}
\caption{Component interface discovery, registration, and matching.
\ding{182} recovers callable interfaces from selected components (with or without export descriptions) into a unified descriptor;
\ding{183} stores descriptors in the  interface registry;
\ding{184} scores candidates against a caller requirement and selects a compatible callee.}
\vspace{-1em}
\label{component_adapter}
\end{figure}

\textbf{\underline{Isolation.}} 
As shown in Tab.~\ref{Tab:sio_compare}, different isolation mechanisms enforce protection by different boundaries, access-control models, and transition operations.
Therefore, the key challenge in designing the isolation abstraction is determining which isolation information should be exposed to upper layers. 
Akira distinguishes between isolation requirements, which describe what protection behavior a component needs, and enforcement states, which describe how a specific mechanism realizes such protection.  
Exposing enforcement states, such as page-table entries, protection keys, or privilege configurations, gives components direct control over low-level protection operations. 
However, these states are mechanism-specific, requiring components to understand different protection models and manage corresponding transitions, which increases complexity and limits heterogeneous composition.
Therefore, Akira exposes only high-level isolation requirements and backend compatibility constraints, while encapsulating mechanism-specific enforcement states.
Specifically, the isolation abstraction consists of three operations:
(1) a lifetime operation, which creates and releases isolation domains;
(2) a transition operation, which enters and leaves protected execution contexts;
and (3) an enforcement operation, which configures and checks access-control policies. 
These requirements capture what components need from isolation mechanisms, while their realization through mechanism-specific enforcement states and execution procedures is encapsulated by isolation backends. 
To make this concrete, Fig.~\ref{fig:iso_abs} contrasts mechanism-specific invocation code with Akira's requirement API.
Without abstraction, MPK and EPT require distinct cross-domain invocation code that directly programs enforcement states.
With Akira, BIU binds and initializes the selected backend through \texttt{fabric\_bridge\_set\_isolation}, while component code issues cross-domain invocations via \texttt{fabric\_gate}; the same three operations---lifetime, transition, and enforcement---are realized inside the selected backend's \texttt{init}/\texttt{enter}/\texttt{leave}, so switching MPK and EPT changes only the binding, not the invocation code. 
\begin{figure*}[h]
\centering
\caption{From mechanism-specific isolation calls to Akira's requirement API.
Left: different isolation exposes different enforcement states.
Right: invocation code expresses requirements only; backends realize
(1)~lifetime, (2)~transition, and (3)~enforcement.}
\vspace{-0.5em} 
\label{fig:iso_abs}
\footnotesize
\begin{minipage}[h]{0.48\textwidth}
\begin{lstlisting}[style=akiraC, title={Before: MPK isolation}]
/* MPK: caller programs PKRU */
pkru = rdpkru();
pkru_set_rw(NET_KEY, &pkru);
wrpkru(pkru);              /* open */
fprintf(stderr, "ERR: %s", errmsg);
pkru_set_none(NET_KEY, &pkru);
wrpkru(pkru);              /* revoke */
\end{lstlisting}

\vspace{0.55em}
\begin{lstlisting}[style=akiraC, title={Before: EPT/VMEPT isolation}]
/* EPT: caller programs VM RPC */
vmept_set_func(ctrl, fprintf, ...);
vmept_wait_for_rpc(ctrl);

/* Switch MPK <-> EPT:
 * rewrite every invocation */
\end{lstlisting}
\end{minipage}
\hfill
\begin{minipage}[h]{0.48\textwidth}
\begin{lstlisting}[style=akiraC, title={After: Akira requirement API}]
/* (1) lifetime: bind+init (once) */
fabric_bridge_set_isolation(br, TYPE, &cfg);
/* TYPE=MPK -> mpk_backend
 * TYPE=EPT -> ept_backend */

/* (2) transition: each call */
fabric_gate(src, fprintf, stderr, "ERR: %s", errmsg);
/* (3) enforcement in init/enter */

/* Same fabric_gate; backends differ
 * only in init/enter/leave */
mpk_backend = {
  mpk_init, mpk_destroy,  /* (1) */
  mpk_enter, mpk_leave};  /* (2)(3) */
ept_backend = {
  ept_init, ept_destroy,  /* (1) */
  ept_enter, ept_leave};  /* (2)(3) */
\end{lstlisting}
\end{minipage}
\end{figure*}
\textbf{\underline{Component.}}
Unlike isolation mechanisms, components continuously evolve in their functional structure, including interfaces, composition patterns, and connection relationships. 
Different implementations of similar functionality may expose different interfaces and invocation conventions, making it difficult to preserve consistent interactions during component replacement. 
The key challenge is not reimplementing components, but deciding which information must be available for one component to invoke another after such a change.
A caller needs the callee's function identity, prototype, parameter layout, ABI, and invocation constraints; it does not need the callee's internal implementation.
If callers depend on that implementation, any replacement forces them to change as well.
If the interface attributes above are missing, the system cannot tell which function may handle a call or how to pass arguments across different ABIs.
Therefore, the AEU extracts these interface attributes, selects a callee that matches them, and rewrites the native call into the fabric's common invocation form, without modifying component implementations, as shown in Fig.~\ref{component_adapter}.

Specifically, before system construction,  Akira performs interface discovery for the selected components. 
In Fig.~\ref{component_adapter} \ding{182}, AEU identifies externally visible component entry points and extracts the interface metadata required for interface matching and adaptation.  
However, components do not always provide explicit interface or ABI-level metadata from which these functionalities and invocation requirements can be read directly. 
Some components provide declarative descriptions (e.g., IDL, manifest, interface tables), while others only expose symbols (e.g., function entry points and binary signatures). 
Akira handles them separately:  
(1) If a component provides explicit export descriptions (e.g., Unikraft’s \texttt{exportsyms.uk}), Akira extracts interface identifiers, ABI versions, function prototypes, and parameter attributes, and generates corresponding registry entries. 
(2) If such declarations are unavailable, Akira recovers interface information in two steps.
Symbol-level discovery scans the component's exported symbols to locate callable entry points when no IDL, manifest, or export table lists them.
ABI/binary analysis then derives each entry's parameter layout from the platform calling convention and binary data rules, because symbol names alone do not specify how arguments are passed. 
When binary information is insufficient (e.g., for dynamically registered callbacks or language-specific abstractions), AEU requires developer-provided interface specifications. 
In Fig.~\ref{component_adapter} \ding{183}, both paths generate a unified interface description template for each callable function, represented as 
$(\mathit{id},\,\mathit{proto},\,\mathit{params},\,\mathit{abi},\,\mathit{ret},\,\mathit{inv})$, which is stored in the interface registry table maintained by CLU (detailed in Sec.~\ref{clu}).

When different component interfaces interact, unlike conventional interface binding mechanisms that rely mainly on ABI equivalence or parameter-type matching, selecting the optimal function from the interface table is fundamentally an interface compatibility problem. 
Simple parameter-type matching methods cannot guarantee functional equivalence, since functions with similar signatures may expose different invocation semantics, ABI or return behaviors. 
Akira introduces an interface matching and verification stage (\ding{184} in Fig.~\ref{component_adapter}). 
Given a caller requirement $r$ and a candidate $c$ from the registry, both represented by $\mathbf{a}(\cdot)$, Akira first applies hard filters on ABI, parameter attributes, and invocation constraints; any candidate that fails is discarded.
For each remaining candidate, it computes a semantic matching score:
\begin{equation}
S(r,c)=\sum_{k\in\{\mathit{id},\mathit{proto},\mathit{ret}\}} w_k\,\delta_k\!\bigl(a_k(r),a_k(c)\bigr),
\qquad \sum_k w_k=1,
\end{equation}
where $a_k(\cdot)$ denotes field $k$ of the interface description $\mathbf{a}(\cdot)$, 
$w_k$ is the weight assigned to field $k$, and $\delta_k=1$ if the corresponding field matches and $0$ otherwise.
Akira then scores every admissible candidate with $S(r,c)$ and selects the candidate with the highest score. 
After a valid target function is selected, Akira automatically transforms the target component’s native ABI and invocation conventions into a unified interface representation.

Overall, through this design, Akira encapsulates different isolation mechanisms and component implementations into reusable system modules with a unified interaction abstraction, enabling flexible composition, replacement, and evolution without requiring system-wide modifications to existing execution paths.

\subsection{Control Logic Unit (CLU)} 
\label{clu} 
AEU provides abstractions for individual modules in Akira, enabling modules to be represented and composed without exposing their internal details. 
However, composing modules requires a dedicated mechanism to coordinate interactions among them. 
Akira therefore introduces the Control Logic Unit (CLU) to manage interaction decisions among modules. 
Building on the module abstractions provided by AEU, CLU separates interaction decisions from module execution paths by introducing an independent control layer between modules. 
CLU maintains global interaction states, including communication relationships, interaction status, and resource availability, and uses these states to determine and coordinate interaction paths through unified control operations. 

\begin{figure}[htpb]
\centerline{\includegraphics[width=0.77\textwidth]{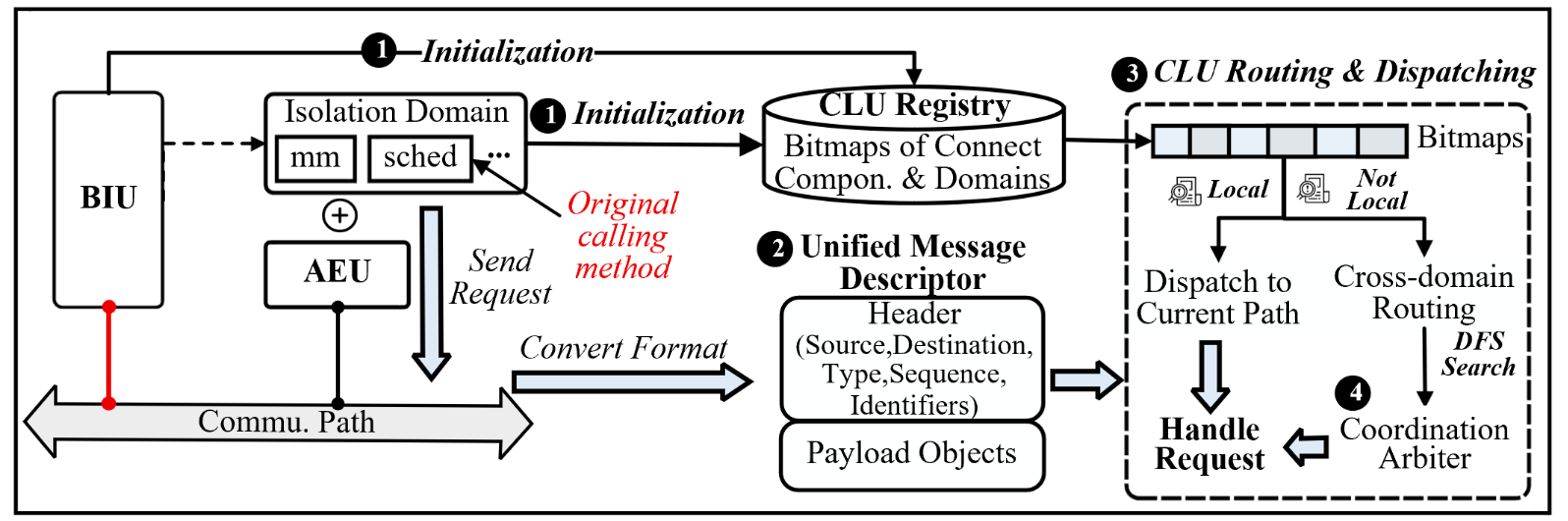}}
\vspace{-0.5em}
\caption{Control Logic Unit (CLU): 
\ding{182}~components register into the CLU bitmap registry at initialization;
\ding{183}~ converts cross-domain invocations into unified message descriptors;
\ding{184}~looks up the destination and dispatches locally;
\ding{185}~ arbiter-mediated contention control.}
\label{controller}
\end{figure} 

\textbf{\textit{Lifetime Management.} }
Each communication fabric in Akira is associated with a dedicated CLU, which maintains bitmaps of connected components and isolation domains, enabling centralized lifecycle management (\ding{182} in Fig.~\ref{controller}). 
During system initialization, components register their interfaces and execution metadata through the AEU adapter into the CLU registry.
This allows the CLU to track active components, their domain assignments, and shared resources. 
When a domain terminates or a component is replaced, the CLU revokes ownership, cleans up capabilities, and removes entries from the registry to prevent dangling references or unauthorized access.

\textbf{\textit{Communication and Routing Management.}} 
To minimize system modification overhead, components within the same domain preserve their original communication mechanisms, such as direct function calls or shared memory. 
For cross-domain interactions, AEU encapsulates different invocation methods into unified message descriptors consisting of structured headers \textit{(source, destination, type, and sequence identifiers)} and payload objects (\ding{183} in Fig.~\ref{controller}).   
Communication fabrics only provide forwarding channels between connected domains, without making routing decisions. 
CLU determines the destination component, selects the appropriate communication path based on maintained bitmaps, and dispatches messages through the corresponding fabric (\ding{184} in Fig.~\ref{controller}). 
During message dispatching, CLU first performs local component lookup through the bitmap to determine whether the destination component resides within the current domain. 
Otherwise, CLU performs cross-domain routing according to system connectivity (bridge upstream/downstream relations and communication fabrics), and adopts a DFS-based search strategy. 
However, as the scale of communication fabrics increases, cross-domain forwarding further introduces path contention and synchronization consistency overheads. 
Akira designs a coordination arbiter inside CLU (\ding{185} in Fig.~\ref{controller}). 
The arbiter maintains a global communication state table to track the runtime states of active cross-domain interactions. 
Unlike registries, the table records dynamic execution states, including active sessions, occupied bridge channels, pending communication requests, shared buffer ownership, and cross-domain synchronization states. 
When a cross-domain request arrives, the arbiter checks whether the target forwarding path or bridge resources are currently occupied by other active sessions. 
Conflicting requests are serialized through pending queues until the corresponding communication resources are released, preventing concurrent forwarding operations from introducing resource contention or inconsistent communication states. 
To avoid exposing global state and compromising isolation, all state updates are mediated by the CLU through coordination interfaces, and each domain can only access communication states associated with its authorized sessions. 
The overhead discussion is in Sec.~\ref{evaluation}.

\textbf{\textit{Interaction Consistency Control.}} 
Because Akira explicitly separates components into isolation domains and connects them through bridges, cross-domain interactions must maintain consistent component-to-domain mappings and communication resource states. 
CLU therefore provides lightweight validation and arbitration mechanisms before establishing a cross-domain interaction. 
For each request, CLU validates the destination component and its assigned isolation domain against the current registry to prevent stale mappings caused by component replacement or domain reconfiguration. 
If the mapping is invalid, CLU rejects the request instead of forwarding it through an outdated bridge path. 
In addition, when multiple requests compete for the same bridge resources, CLU arbitrates access to communication channels and shared buffers before establishing the interaction path, ensuring that resource allocation remains consistent across domains. 
\subsection{Bridge-based Isolation Unit (BIU)} 
\label{biu} 
\begin{figure}[htpb]
\centerline{\includegraphics[width=\textwidth]{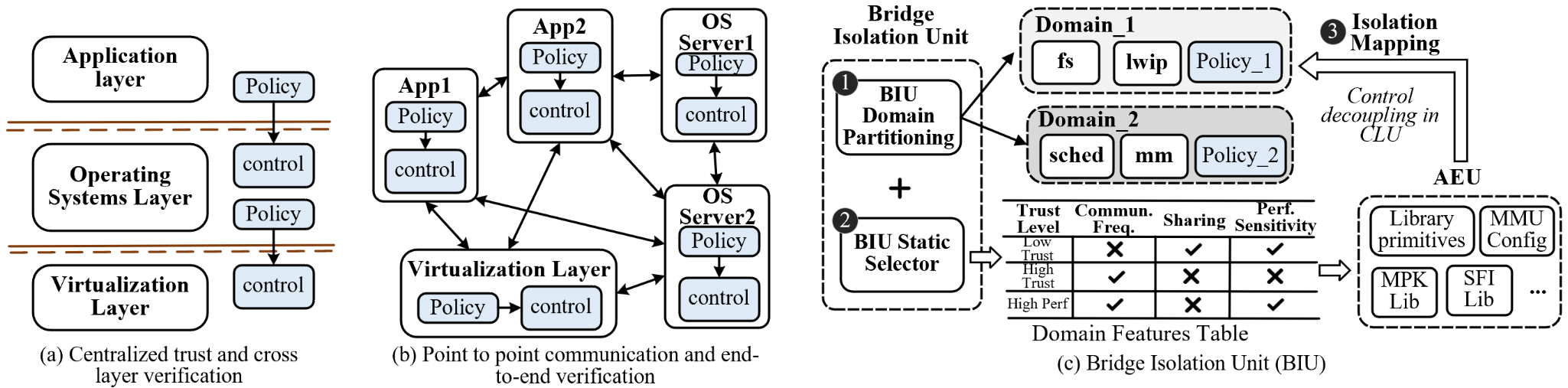}}
\vspace{-0.5em}
\caption{Isolation organization: (a)~centralized cross-layer trust; (b)~point-to-point end-to-end verification; (c)~BIU domain partitioning, static backend selection, and AEU isolation mapping.} 
\vspace{-1em}
\label{bridge1}
\end{figure}

With the CLU establishing module interactions, Akira can flexibly coordinate communication between modules. 
However, interactions between modules may require different isolation guarantees (insight\#3). Thus, Akira further introduces the BIU to manage isolation requirements, separating protection management from interaction control.  
Existing approaches either rely on centralized isolation management with limited flexibility (Fig.~\ref{bridge1} (a)) or distribute protection management across components with increased management overhead (Fig.~\ref{bridge1} (b)). 
Akira introduces the BIU to construct configurable isolation boundaries across module interactions. 
The BIU groups modules with similar protection requirements and communication characteristics into isolation domains, enabling each domain to adopt an appropriate isolation mechanism (Fig.~\ref{bridge1} (c)).

Specifically, 
each isolation domain is then assigned an appropriate isolation mechanism through AEUs, enabling different protection boundaries to coexist within the same system. 
This design enables Akira to select appropriate isolation boundaries and mechanisms for different module requirements, avoiding the performance costs of unnecessary isolation while preserving required protection guarantees. 
Achieving such configurable isolation requires the BIU to make two related decisions: 
\textbf{(1) domain construction:} determining which components should be placed in the same isolation domain (\ding{182} in Fig.~\ref{bridge1} (c)) based on their protection requirements and communication dependencies, balancing isolation boundaries and interaction costs.  
\textbf{(2) backend assignment:} determining which isolation mechanism provided by AEUs should be assigned to each constructed domain (\ding{183}-\ding{184} in Fig.~\ref{bridge1} (c)) according to its required protection level.

Although components provide functional modularity, component boundaries do not necessarily correspond to isolation boundaries. 
Separating all components into independent isolation domains provides strong protection boundaries, but introduces additional cross-domain transitions when frequently communicating components are isolated from each other. 
Conversely, placing communicating components into the same domain reduces transition overhead, but may not be feasible when components have incompatible isolation requirements or backend constraints. 
Therefore, BIU constructs isolation domains by balancing communication efficiency and isolation constraints.


\begin{wrapfigure}{r}{0.54\columnwidth}
\vspace{0pt}

\begin{algorithm}[H]
\caption{BIU Domain Partitioning}
\small
\label{alg:biu_partition}
\KwIn{Component set $C$; communication matrix $F$; isolation backend candidates $A(c)$; merge threshold $\theta$}
\KwOut{Isolation domain set $D$}

$D \leftarrow \emptyset$\;
\SetKwFor{ForEach}{foreach}{do \CommentSty{/* Init Domains */}}{}
\ForEach{$c \in C$}{
    $d \leftarrow \{c\}$\;
    $A(d) \leftarrow A(c)$\;
    $D \leftarrow D \cup \{d\}$\;
}
\SetKwFor{ForEach}{foreach}{do}{}

\ForEach{domain pair $(d_i,d_j)$}{
    $intensity \leftarrow \sum_{c \in d_i}\sum_{c' \in d_j} F[c,c']$\;
    \textbf{BIU}.Enqueue($d_i,d_j,intensity$)\;
}

\While{\textbf{BIU}.HasCandidate()}{

    $(d_i,d_j,intensity) \leftarrow$ \textbf{BIU}.SelectCandidate()\;
    \textbf{BIU}.Dequeue($d_i,d_j,intensity$)\;

    \If{\textbf{BIU}.Compatible($d_i,d_j$)}{

        $d_{new} \leftarrow$ \textbf{BIU}.Merge($d_i,d_j$)\;

        $A(d_{new}) \leftarrow A(d_i)\cap A(d_j)$\;

        \textbf{BIU}.UpdateCandidates($d_{new}$)\;

        \textbf{BIU}.AssignBackend($d_{new}$)\;
    }
}

\Return{$D$}\;
\end{algorithm}

\vspace{-8pt}
\end{wrapfigure}
As shown in Algo.~\ref{alg:biu_partition}, 
BIU constructs isolation domains before deployment based on component isolation constraints and workload profiling results. 
For each component, developers specify its isolation requirements and backend compatibility constraints in a YAML configuration file. 
These specifications define the feasible isolation configurations for each component and constrain the domain compositions explored by BIU, rather than explicitly defining the final isolation domains. 
Before deployment, BIU performs workload profiling to characterize component interactions for static isolation domain construction. 
Since isolation domains are determined at deployment time, BIU analyzes the structural dependencies among components rather than transient runtime states. 
During profiling, Akira executes representative workloads and records component invocations through the Fabric interface. 
For a workload $w_k$, BIU records the number of invocations from component $c_i$ to component $c_j$ as $N_{ij}^{(k)}$ and defines the invocation rate as:
$F^{(k)}[c_i,c_j]=N_{ij}^{(k)}/T_k$,
where $T_k$ denotes the workload execution duration. 
Each entry $F^{(k)}[c_i,c_j]$ represents the observed invocation rate from component $c_i$ to component $c_j$ during workload $w_k$. 
For applications containing multiple workloads, BIU aggregates the invocation rates according to workload weights:
\begin{equation}
F[c_i,c_j] =
\begin{cases}
\sum_{k=1}^{K} p_k F^{(k)}[c_i,c_j],
& i \neq j,\\
0, & i = j,
\end{cases}
\end{equation}
where $p_k$ denotes the weight of workload $w_k$, with $\sum_{k=1}^{K}p_k=1$.
The aggregated matrix $F$ captures component interaction patterns used by BIU to identify components that may benefit from co-location. 
Using the aggregated matrix $F$ and component isolation constraints, BIU constructs isolation domains through a constraint-preserving merging procedure (\ding{182} in Fig.~\ref{bridge1}(c)). 
Starting from an initial domain composition, BIU merges components only when the resulting composition satisfies isolation and backend compatibility constraints while reducing unnecessary cross-domain transitions. 
Thus, workload profiling guides domain consolidation, whereas developer-provided constraints determine which merges are allowed.



After constructing isolation domains, BIU assigns an isolation backend to realize each domain boundary (\ding{183} and \ding{184} in Fig.~\ref{bridge1} (c)). 
AEUs provide heterogeneous isolation mechanisms with different protection capabilities and performance overheads, resulting in a large design space for backend selection. 
To simplify this selection, BIU first classifies these mechanisms according to the protection boundaries they establish (Tab.~\ref{Tab:sio_compare}), and then selects a suitable backend within the corresponding category based on domain requirements. 
The supported protection boundaries include:
(1) shared execution without additional boundaries (BD0);
(2) shared execution with restricted resource access (BD1); and
(3) separated execution with independent privilege contexts (BD2). 
This hierarchical selection avoids directly exploring all available isolation mechanisms and captures the protection-performance trade-offs across different isolation levels, enabling BIU to choose an isolation backend that satisfies required protection guarantees while minimizing unnecessary performance overhead.


\textbf{\textit{Bridge Domain 0 (BD0): Shared-Execution Backend. }}
When BIU selects BD0, it places the involved components into the same trusted execution domain without attaching any additional isolation backend.
AEU preserves the existing execution context and memory-access configuration rather than configuring additional isolation metadata or boundary transitions. 
CLU routes interactions by the native communication path without invoking isolation entry or exit operations.
Since no additional protection checks are introduced, BD0 provides the lowest communication overhead and is suitable for mutually trusted components with frequent interactions.

\textbf{\textit{Bridge Domain 1 (BD1): Resource-Access Isolation Backend. }}
BD1 realizes resource-access isolation by restricting how components access each other's resources while allowing them to remain within the same execution context. 
When BIU selects BD1 for an isolation domain, it binds the domain to a resource-access backend and initializes the corresponding protection metadata, such as MPK keys, MPU regions, MTE tags, or SFI access policies. 
During component interaction, CLU invokes the backend-specific enforcement operations through unified isolation interfaces, where the backend applies mechanisms such as permission switching, region reconfiguration, or runtime access checks. 
These mechanisms restrict unauthorized accesses to shared resources, such as memory regions, without requiring separate address spaces or execution contexts.

\textbf{\textit{Bridge Domain 2 (BD2): Execution-Context Isolation Backend. }} 
When BIU selects BD2, it binds the isolation domain to an execution-context backend and allocates an independent execution context, such as a separate page-table hierarchy, virtual machine context, privilege context, or trusted execution environment. 
AEU continues to register components and encapsulate their interfaces through the common abstraction, but does not construct page tables, configure privilege states, or perform execution-context transitions. 
CLU invokes the backend-specific isolation operations during cross-domain interactions, where the backend switches to the target execution context through mechanisms such as address-space switching, virtualization transitions, or secure-world switching before dispatching the request. 
Since each component executes within an independent execution context, BD2 provides strong containment for components requiring strict isolation.

Overall, this design provides configurable isolation, adapting the isolation granularity to the performance and security requirements of different components (overhead analyzed in Sec.~\ref{evaluation}).


\section{Case Study: Using Akira}
\label{case_stduy} 
We choose InfiniTime~\cite{Github-InfiniTime}, an open-source FreeRTOS-based smartwatch firmware with multiple functional components executing within a shared protection domain. Using InfiniTime as a target system, this case study demonstrates how Akira adapts an existing OS without redesigning its components, including why reconfiguration is needed, what remains unchanged, and how AEU, CLU, and BIU manage component encapsulation, communication, and isolation through configuration. 

\subsection{Baseline System} 
In the original InfiniTime design, all components are statically linked with FreeRTOS and execute within a single address space.
The system contains several classes of components:

\begin{itemize}
\item \textbf{Core system components}: FreeRTOS kernel services, task management, timers, and system coordination logic provide scheduling and lifecycle management.
\item \textbf{Device service  components}: device drivers and controllers manage peripherals including LCD, touch sensors, IMU, battery, and motor devices. 
\item \textbf{Connectivity components}: the Bluetooth stack, GATT services, DFU, and file transfer services process external communication.
\item \textbf{Application components}: UI frameworks, watch faces, and user applications provide extensibility but may contain less trusted code.
\end{itemize}

Although these components have different security requirements, the original system does not enforce isolation between them.
Components communicate through direct C/C++ function calls and FreeRTOS queues.
For example, a UI component can directly invoke the motor controller, and a Bluetooth service can send messages to system components through shared queues (Listing~\ref{lst:orig_infinitime}).
This design provides efficient communication, but a faulty or compromised component may access unrelated resources because all components share the same protection domain.

\noindent
\begin{minipage}[t]{0.48\columnwidth}
\begin{lstlisting}[
language=C,
basicstyle=\ttfamily\scriptsize,
frame=single,
breaklines=true,
columns=fullflexible,
keepspaces=true,
numbers=left,
numberstyle=\tiny\color{gray},
numbersep=5pt,
escapeinside={@}{@},
caption={Original InfiniTime interaction.},
label={lst:orig_infinitime}
]
QueueHandle_t systemTasksMsgQueue;
void SystemTask::Work() {
    Messages msg;
    @\textbf{batteryController.MeasureVoltage();}@
    xQueueReceive(systemTasksMsgQueue, &msg, waitTime);
    if (msg == Messages::OnNewNotification)
        displayApp.PushMessage(NewNotification);
    if (msg == Messages::MeasureBatteryTimerExpired)
        @\textbf{batteryController.MeasureVoltage();}@
}
void QuickSettings::OnButtonEvent() { /* UI, same AS */
    @\textbf{motorController.RunForDuration(35);}@
}
void @\textbf{SystemTask::PushMessage}@(Messages msg) {
    xQueueSend(systemTasksMsgQueue, &msg, portMAX_DELAY);
}
\end{lstlisting}
\end{minipage}
\hfill
\begin{minipage}[t]{0.48\columnwidth}
\begin{lstlisting}[
language=C,
basicstyle=\ttfamily\scriptsize,
frame=single,
breaklines=true,
columns=fullflexible,
keepspaces=true,
numbers=left,
numberstyle=\tiny\color{gray},
numbersep=5pt,
escapeinside={@}{@},
caption={AEU encapsulation of Listing~\ref{lst:orig_infinitime}.},
label={lst:aeu_wrap}
]
static void @\textbf{Handler\_Battery\_MeasureVol}@() {
  batteryController.MeasureVoltage();
}
static void @\textbf{Handler\_Motor\_Run}@() {
  motorController.RunForDuration(d);
}
static void @\textbf{Handler\_SystemTask\_PushMessage}@() {
  systemTask.PushMessage(msg);
}
void RegisterInfiniTime(SoftBusComponent* c) {
  auto& ctl = c->getComponentController();
  ctl.registerFunction(Endpoints::kBatteryMeasure,
                       Handler_Battery_MeasureVol);
  ctl.registerFunction(Endpoints::kMotorRun,
                       Handler_Motor_Run);
  ctl.registerFunction(
      Endpoints::kSystemTaskPushMessage,
      Handler_SystemTask_PushMessage);
}
\end{lstlisting}
\end{minipage}

\subsection{Applying Akira}
Akira enables developers to adapt InfiniTime through configuration rather than modifying component implementations.
The migration requires only two decisions:
(1) identifying existing components and their interfaces, and
(2) specifying the desired component grouping, communication paths, and isolation mechanisms. 

\textbf{Step~1: Encapsulate existing components through the AEU.}
Akira first uses the AEU to wrap the existing InfiniTime and FreeRTOS modules as Akira components. 
In the original InfiniTime, components invoke one another through native C/C++ calls and FreeRTOS queues within a single address space (Listing~\ref{lst:orig_infinitime}, lines~4, 11, and~13).
To migrate these modules onto Akira, the developer does not rewrite their implementations; instead, the developer only designates which existing entry points should be exposed as component interfaces.

Because FreeRTOS and InfiniTime do not ship declarative export descriptions such as IDL files, manifests, or Unikraft-style \texttt{exportsyms.uk} tables, AEU recovers their interfaces along the symbol-based path of Akira's interface discovery (Sec.~\ref{aeu}).
Specifically, symbol-level discovery scans the selected modules' globally visible C/C++ symbols to locate callable entry points; ABI/binary analysis then derives each entry's parameter layout from the ARM Cortex-M calling convention and binary data rules.
From these  descriptors, AEU automatically generates an encapsulation
layer and registers it in the interface registry, while leaving the method body unchanged (Listing~\ref{lst:aeu_wrap}).

\textbf{Step~2: Configure domains and isolation backends with the BIU.}
The developer specifies component-level isolation requirements, acceptable backends, and the merge threshold in \texttt{manual.yaml} (Listing~\ref{lst:akira_constraints}); the \textbf{bold} values are filled in manually.
The input does not specify domain or bridges.
Together with the workload profile, BIU uses these constraints to partition the components and assign a compatible backend to each resulting domain.
It then generates \texttt{domain.yaml}, a deployment manifest containing the derived domains, backends, and bridges (Listing~\ref{lst:akira_template}).

\definecolor{isoBD0}{RGB}{0,102,153}
\definecolor{isoBD1}{RGB}{0,128,105}
\definecolor{isoBD2}{RGB}{176,78,28}
\definecolor{isoThr}{RGB}{153,40,40}

\noindent
\begin{minipage}[t]{0.48\columnwidth}
\begin{lstlisting}[
basicstyle=\ttfamily\scriptsize,
frame=single,
breaklines=true,
columns=fullflexible,
numbers=left,
numberstyle=\tiny\color{gray},
numbersep=5pt,
escapeinside={@}{@},
caption={Developer constraints (\texttt{manual.yaml}).},
label={lst:akira_constraints}
]
merge_threshold: @\textcolor{isoThr}{\textbf{500}}@  # invocations/s
components:
  freertos_tasks:  {level: @\textcolor{isoBD0}{\textbf{BD0}}@, backends: @\textcolor{isoBD0}{\textbf{[native]}}@}
  freertos_timers: {level: @\textcolor{isoBD0}{\textbf{BD0}}@, backends: @\textcolor{isoBD0}{\textbf{[native]}}@}
  freertos_queues: {level: @\textcolor{isoBD0}{\textbf{BD0}}@, backends: @\textcolor{isoBD0}{\textbf{[native]}}@}
  spi_master: {level: @\textcolor{isoBD1}{\textbf{BD1}}@, backends: @\textcolor{isoBD1}{\textbf{[mpu]}}@}
  battery_controller: {level: @\textcolor{isoBD1}{\textbf{BD1}}@, backends: @\textcolor{isoBD1}{\textbf{[mpu]}}@}
  motor_controller: {level: @\textcolor{isoBD1}{\textbf{BD1}}@, backends: @\textcolor{isoBD1}{\textbf{[mpu]}}@}
  nimble_host: {level: @\textcolor{isoBD1}{\textbf{BD1}}@, backends: @\textcolor{isoBD1}{\textbf{[mpu]}}@}
  dfu_service: {level: @\textcolor{isoBD1}{\textbf{BD1}}@, backends: @\textcolor{isoBD1}{\textbf{[mpu]}}@}
  notification_manager: {level: @\textcolor{isoBD1}{\textbf{BD1}}@, backends: @\textcolor{isoBD1}{\textbf{[mpu]}}@}
  display_app: {level: @\textcolor{isoBD2}{\textbf{BD2}}@, backends: @\textcolor{isoBD2}{\textbf{[priv\_switch]}}@}
  user_apps: {level: @\textcolor{isoBD2}{\textbf{BD2}}@, backends: @\textcolor{isoBD2}{\textbf{[priv\_switch]}}@}
\end{lstlisting}
\end{minipage}
\hfill
\begin{minipage}[t]{0.48\columnwidth}
\begin{lstlisting}[
basicstyle=\ttfamily\scriptsize,
frame=single,
breaklines=true,
columns=fullflexible,
numbers=left,
numberstyle=\tiny\color{gray},
numbersep=5pt,
escapeinside={@}{@},
caption={BIU deployment (\texttt{domain.yaml}).},
label={lst:akira_template}
]
domains:
- name: kernel_domain
  isolation: @\textcolor{isoBD0}{\textbf{native}}@            # BIU @\textcolor{isoBD0}{\textbf{BD0}}@
  comps: [freertos_tasks, freertos_timers, freertos_queues]
- name: driver_domain
  isolation: @\textcolor{isoBD1}{\textbf{mpu}}@               # BIU @\textcolor{isoBD1}{\textbf{BD1}}@
  comps: [spi_master, battery_controller, motor_controller]
- name: connectivity_domain
  isolation: @\textcolor{isoBD1}{\textbf{mpu}}@               # BIU @\textcolor{isoBD1}{\textbf{BD1}}@
  comps: [nimble_host, dfu_service, notification_manager]
- name: app_domain
  isolation: @\textcolor{isoBD2}{\textbf{priv\_switch}}@       # BIU @\textcolor{isoBD2}{\textbf{BD2}}@
  comps: [display_app, user_apps]
bridges:
- {from: kernel_domain, to: driver_domain}
- {from: driver_domain, to: connectivity_domain}
- {from: connectivity_domain, to: app_domain}
\end{lstlisting}
\end{minipage}

From the developer constraints and the profiled communication matrix, Akira's BIU partitioning procedure
(Algo.~\ref{alg:biu_partition}) automatically derives domain, grouping InfiniTime into four domains: kernel, driver, connectivity, and application.
It records the resulting membership as the \texttt{comps} entries in \texttt{domain.yaml}, assigns each domain a backend from the intersection of its components' admissible backend sets, and derives bridges for the required cross-domain interactions.

The kernel domain uses BD0 (\texttt{native}): no additional MPU region or privilege transition is attached, so FreeRTOS scheduling, tickless idle, and intra-domain queues remain on their original privileged execution path, minimizing runtime overhead. 
The driver and connectivity domains use BD1 (\texttt{mpu}): BIU configures ARM MPU regions to restrict each domain's accessible memory and peripherals, preserving efficient local I/O operations while preventing a faulty driver or untrusted BLE/DFU handler from directly corrupting kernel stacks or issuing arbitrary GPIO accesses. 
The application domain uses BD2 (\texttt{priv\_switch}): BIU places \texttt{DisplayApp} and user applications in Cortex-M unprivileged USER mode and mediates privileged operations through configured domain transitions, enforcing least privilege for the largest and least-trusted software surface.

\textbf{Step~3: Generate communication paths and enforce isolation. }
Given the configuration above, Akira automatically generates the required communication paths and isolation setup at build time (Listings~\ref{lst:orig_cross} and~\ref{lst:cross_invoke}).
CLU connects domains through the declared bridges, and BIU applies the selected backends to those domains.
Calls inside the same domain keep their original InfiniTime paths; only cross-domain calls are redirected through Akira.
As a result, InfiniTime retains its original functionality while gaining configurable isolation.
No kernel, driver, Bluetooth, or application source needs to be redesigned: Akira separates component implementations from system-level coordination and isolation decisions.
Therefore, the same components can support different deployments by changing only the configuration.

\noindent
\begin{minipage}[t]{0.42\columnwidth}
\begin{lstlisting}[
language=C,
basicstyle=\ttfamily\scriptsize,
frame=single,
breaklines=true,
columns=fullflexible,
keepspaces=true,
numbers=left,
numberstyle=\tiny\color{gray},
numbersep=5pt,
caption={Original cross-module calls.},
label={lst:orig_cross}
]
/* QuickSettings.cpp */
void OnButtonEvent() {
    motorController.RunForDuration(35);
}
/* SystemTask.cpp */
case MeasureBatteryTimerExpired:
    batteryController.MeasureVoltage();
/* AlertNotificationClient */
systemTask.PushMessage(
    Messages::OnNewNotification);
\end{lstlisting}
\end{minipage}
\hfill
\begin{minipage}[t]{0.52\columnwidth}
\begin{lstlisting}[
language=C,
basicstyle=\ttfamily\scriptsize,
frame=single,
breaklines=true,
columns=fullflexible,
keepspaces=true,
numbers=left,
numberstyle=\tiny\color{gray},
numbersep=5pt,
caption={CLU invocation (app domain to driver domain).},
label={lst:cross_invoke}
]
fabric_gate(MotorController, "Motor_Run");
fabric_gate(BatteryController,
            "Battery_MeasureVol");
void fabric_gate(CompId dst, const char *fn) {
    SoftBusMessage msg(DisplayApp, dst, fn);
    auto *ctrl = routeRequest(src, &msg);
    Bridge *br = ctrl->getUplinkBridge();
    if (br->dispatchUnprivileged())
        portSWITCH_TO_USER_MODE();
    ProcessMessage(ctrl, &msg);
}
\end{lstlisting}
\end{minipage}

\section{Evaluation}
\label{evaluation}
Based on Akira's design, we select two representative OSs, FreeRTOS~\cite{barry2008freertos} and Unikraft~\cite{kuenzer2021unikraft}, as component sources to construct customized OS configurations for different application scenarios. 
FreeRTOS adopts a monolithic architecture where functionalities are organized as kernel modules but remain tightly coupled, providing a representative case to evaluate Akira's ability to decouple and reconstruct existing OS components.
In contrast, Unikraft adopts a library OS architecture that explicitly decomposes OS functionalities into configurable components, enabling the evaluation of Akira's capability in composing and managing highly modular systems. 
By rebuilding OS instances through Akira using components from both systems, we evaluate its effectiveness in enabling agile OS specialization while maintaining performance and scalability across different system scales. 
The reconstructed instances are functionally equivalent to their original counterparts, as Akira only changes component organization and interaction mechanisms without altering OS functionalities.

\noindent
\underline{\textbf{Basic Setup.}}
We evaluate FreeRTOS-based systems on an LPC55S69 board with dual 150 MHz Arm Cortex-M33 cores, 640 KB flash, and 320 KB SRAM, and compile the firmware using the Arm GNU Toolchain. 
We run the Unikraft-based experiment on a server equipped with two AMD EPYC 7763 processors (256 logical CPU cores in total) and 503 GiB memory. 
All measurements are repeated at least 10 times, and we report the average results to ensure statistical reliability. 

\noindent
\underline{\textbf{Workloads.}}
We evaluate Akira across three representative benchmark suites to validate its effectiveness in diverse computing scenarios: 
the micro-benchmark for cross-domain communication overhead under isolation transitions (e.g., empty round-trips and multi-hop domain chains),
the Embench~\cite{embench2020} for resource-constrained embedded scenarios dominated by OS primitive latency and privileged kernel-path operations, and 
Nginx~\cite{nginx} and Redis~\cite{redis} for real-world server workloads dominated by multi-component interaction under reconfigurable isolation.

\noindent
\underline{\textbf{Baseline.} }
We select representative systems from two categories for comparison. 
The traditional OS baseline includes systems that rely on fixed protection boundaries, such as user/kernel separation or isolated execution domains. 
They can be viewed as a baseline consisting of fixed isolation domains and used to evaluate the performance cost introduced by Akira's flexible component composition and communication mechanisms. 
Representative examples include microkernel systems (e.g., seL4~\cite{klein2009sel4} and Fiasco.OC~\cite{smejkal2015transactional}), and hybrid kernels (e.g., Zircon~\cite{zircon}). 
The customizable OS baseline includes systems that relax fixed OS structures and explore more flexible system organizations or execution environments. 
Representative examples include FlexOS~\cite{lefeuvre2022flexos}, UnderBridge~\cite{gu2020harmonizing}, and SkyBridge~\cite{mi2019skybridge}. 
These systems are selected to evaluate Akira's advantages in agile specialization, component organization, and scalability. 

\subsection{Agility Exploration: FreeRTOS and Unikraft} 
To evaluate Akira's ability to reconstruct OSs from components with different levels of modularity, we conduct experiments on FreeRTOS and Unikraft, which represent distinct points in the OS design space.  
Using components from these two systems, Akira constructs OS configurations with different isolation backends. 
For FreeRTOS-based systems, Akira adopts privilege switching and MPU~\cite{armv8mmpu} to provide lightweight isolation for resource-constrained embedded environments. 
For Unikraft-based systems, Akira leverages hardware-assisted mechanisms, including Intel Memory Protection Keys (MPK)~\cite{corbet2015mpk} and Extended Page Tables (EPT)~\cite{intel2015mpx}, to provide configurable isolation for modular server-oriented systems. 
These experiments evaluate whether Akira can enable agile OS specialization across architectures with different modularity and isolation mechanisms. 

\begin{figure}[htpb]
\centerline{\includegraphics[width=0.94\textwidth]{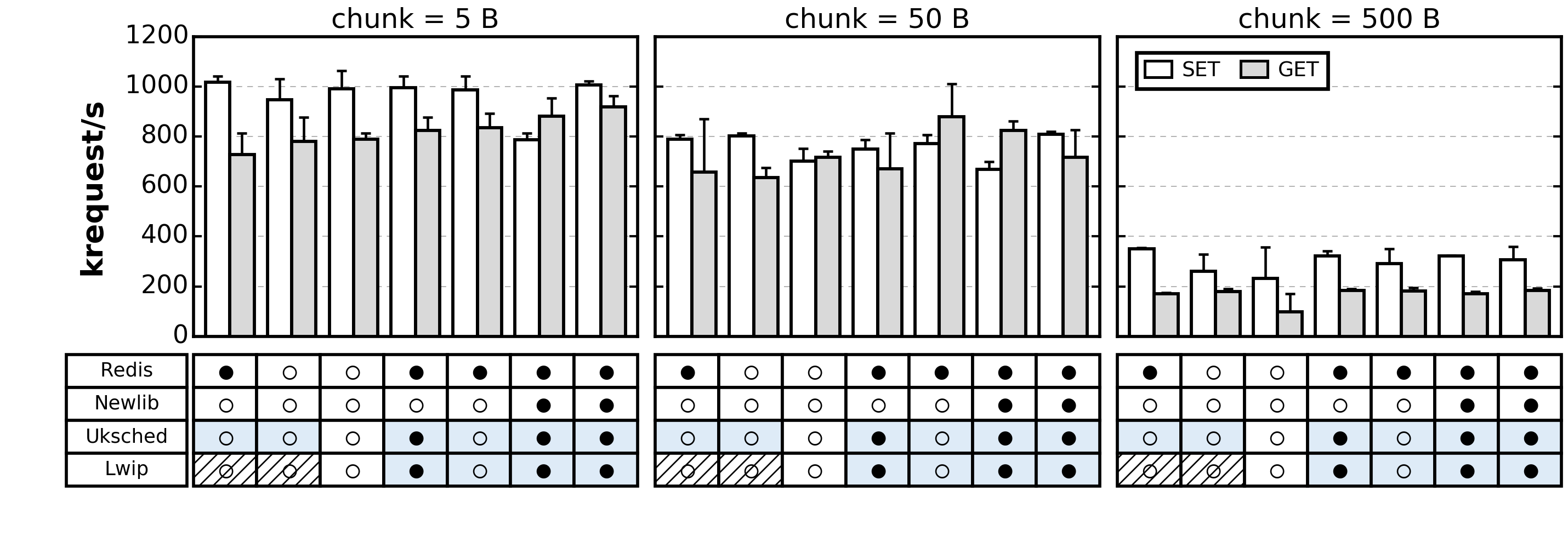}}
\vspace{-2.0em}
\caption{Redis performance for a range of configurations. 
Components are on the left. Software hardening can be enabled [\textbullet] or disabled [$\circ$] for each component. 
The white, blue, and hatched backgrounds denote different isolation compartments, each enforced by a separate MPK protection domain. } 
\vspace{-1.0em}
\label{redis}
\end{figure} 

\begin{figure}[htpb]
    \centering
    \begin{minipage}{0.48\linewidth}
        \centering
        \includegraphics[width=0.95\linewidth]{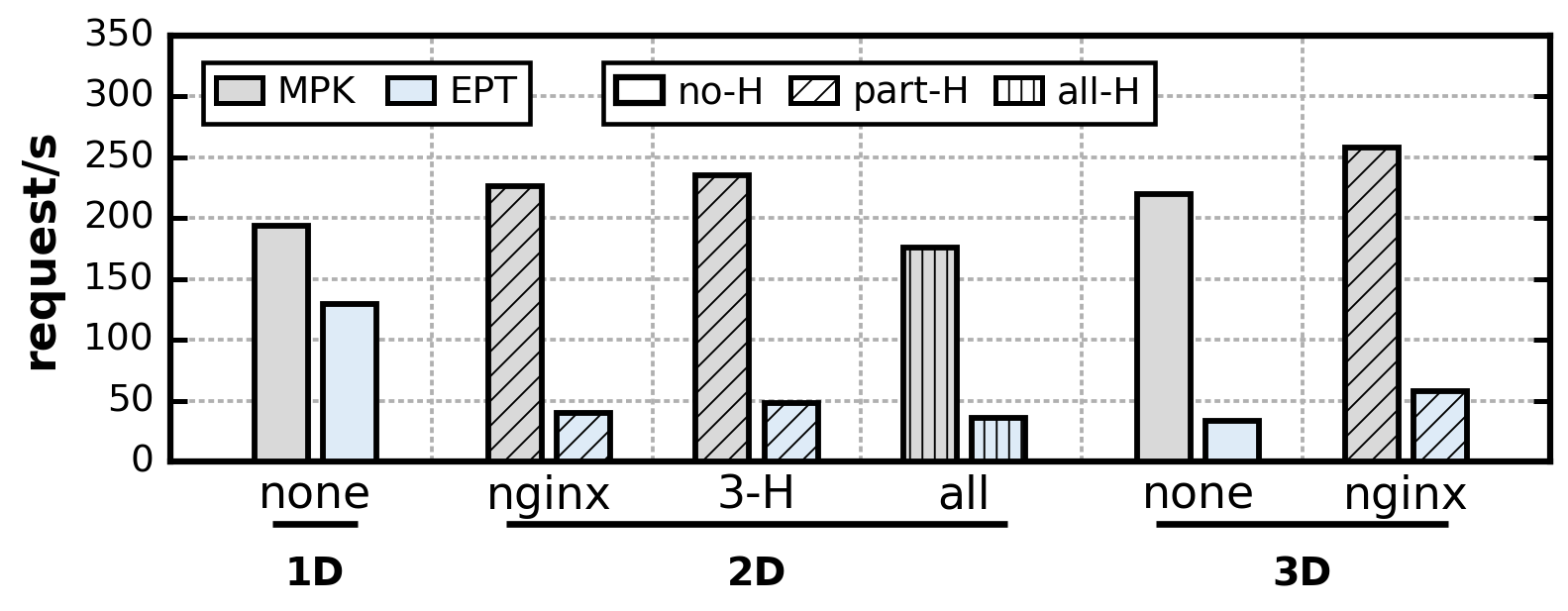}
        \vspace{-10pt}
        \caption{Nginx throughput under MPK and EPT. \texttt{1D}/\texttt{2D}/\texttt{3D}: domain count. Labels: hardening scope---\texttt{none}; \texttt{nginx} (app only); \texttt{3-H} (\texttt{nginx}+\texttt{uksched}+\texttt{lwip}); \texttt{all} (four libraries). \texttt{nginx} and \texttt{3-H} use the same 2D split and differ only in hardening.}
        \label{nginx_mpk}
    \end{minipage}\hfill
    \begin{minipage}{0.5\linewidth}
        \centering
        \includegraphics[width=0.97\linewidth]{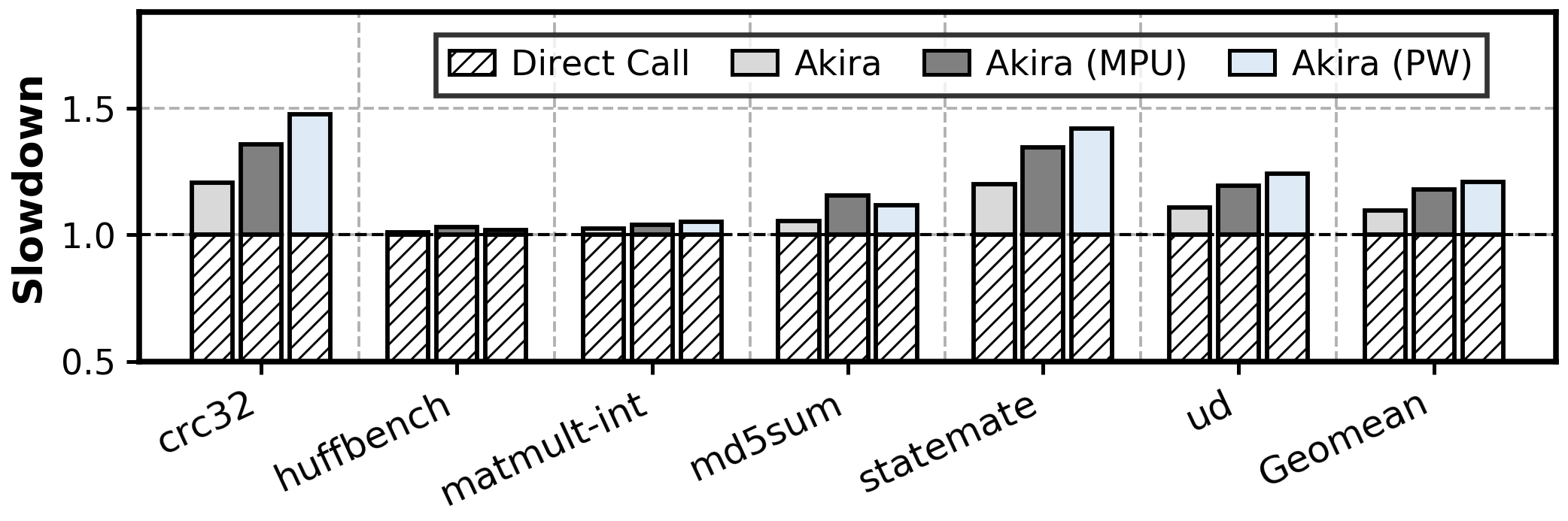}
        \vspace{-10pt}
        \caption{Embench performance of Akira on FreeRTOS under different isolation backends. Direct Call (baseline); Akira (no isolation); Akira (MPU); and
Akira (PW) with privilege-level switching.} 
        \label{agile_freertos}
    \end{minipage}
    \vspace{-10pt}
\end{figure}

\begin{table}[h]
\begin{minipage}[t]{0.47\linewidth}
\centering
\vspace{-7.5em}
\caption{Code modification for evolving isolation on FreeRTOS (KLOC). Direct FreeRTOS changes are intrusive; Akira changes are incremental.}
\label{tab:freertos_agility_kloc}
\footnotesize
\setlength{\tabcolsep}{3pt}
\begin{tabular}{@{}lcc@{}}
\toprule
Change & FreeRTOS & Akira \\
\midrule
\multicolumn{3}{@{}l@{}}{\textbf{\emph{I. One-time framework (written once, reused)}}} \\
Fabric & --- & 0.49 \\
Bridge & --- & 0.24 \\
Controller & --- & 1.42 \\
\midrule
\multicolumn{3}{@{}l@{}}{\textbf{\emph{II. Bind / switch isolation backend}}} \\
MPU isolation & 1.45 & 0.23 \\
Privilege switch & 1.30 & 0.32 \\
\textit{(Location)} & \textit{Source Code} & \textit{config.yaml} \\
\textbf{Average} & \textbf{1.38} & \textbf{0.28} \\
\bottomrule
\end{tabular}
\end{minipage}
\begin{minipage}[h]{0.47\linewidth}
\centering
\caption{Modification  when evolving FreeRTOS (\%).
Without Akira, adding MPU/privilege/I/O/TCP triggers chained edits across core files. Using Akira, only configuration updates are required.}
\label{tab:freertos_modified_pct}
\footnotesize
\setlength{\tabcolsep}{2pt}

\begin{tabular}{@{}c|lccccc@{}}
\toprule
& & \multicolumn{4}{c}{\textbf{Direct FreeRTOS (\%)}} &\textbf{ Akira }\\

\textbf{Location} & \textbf{Metric} & \emph{+MPU} & \emph{+Priv.} & \emph{+I/O} & \emph{+TCP} & \emph{(All)} \\
\midrule
\multirow{4}{*}{source code}
            & Layers       & 7.32  & 9.61  & 9.67  & 11.33 & $\approx$0 \\
            & Core files   & 19.04 & 21.02 & 20.87 & 22.66 & $\approx$0 \\
            & Chained      & 52.30 & 45.35 & 46.31 & 39.06 & $\approx$0 \\
            & Single-layer & 21.34 & 24.02 & 23.16 & 26.95 & $\approx$0 \\
\midrule
config.yaml & Config.      & ---   & ---   & ---   & ---   & \ding{52} \\
\bottomrule
\end{tabular}
\end{minipage}
\end{table} 

\noindent
\underline{\textbf{Unikraft.}} 
We evaluate Akira on Unikraft using two real-world applications: Redis~\cite{redis} (a database server) and Nginx~\cite{nginx} (an HTTP server). 
We instantiate Akira with two isolation backends, MPK and EPT, and explore different configurations by varying the number of isolation domains ($1$--$3$), the partitioning of components across domains (TCP/IP stack, libc, scheduler, and application), and the software hardening mechanisms enabled for each domain (UBSan and KASan). 
These configurations represent different points in the isolation-performance design space. 

\textbf{\textit{Redis.} }
The results are shown in Fig.~\ref{redis}, which evaluates Redis GET/SET throughput under different workload scales (through different chunk configurations) using Akira with the MPK isolation backend.  
Overall, Akira provides flexible safety configurations, spanning a wide performance range from 97.7 to 1017.7~kreq/s, with up to a 10.4$\times$ throughput variation across different protection settings. 
This performance variation reflects the trade-off between protection and performance across different safety configurations. 
However, performance does not degrade proportionally with the number of isolation domains, as the impact depends on component interactions. 
For example, separating \texttt{lwip} from a shared \texttt{lwip}/\texttt{uksched} domain reduces GET throughput by 13.0\%, due to increased cross-domain communication. 
Therefore, Akira's partitioning design (Algo.~\ref{alg:biu_partition}) considers interaction patterns and co-locates highly interacting components.  

\textbf{\textit{Nginx.} }
We evaluate Nginx HTTP throughput under MPK and EPT backends across different domain organizations and software hardening configurations (Fig.~\ref{nginx_mpk}). 
We observe that Akira enables a wide range of system configurations, with different isolation and hardening choices causing up to a 7.8$\times$ throughput variation. 
This variation is affected by the underlying isolation mechanism, as EPT-based isolation consistently introduces higher overhead than MPK-based isolation across configurations. 
The impact of a given partitioning strategy also depends on workload characteristics. 
For example, the same \texttt{lwip}-scheduler separation causes only a 3.1\% throughput loss for Nginx but a larger 13\% degradation for Redis GET due to different interaction patterns. 

Overall, these results demonstrate that Akira exposes a wide security-performance configuration space through configurable component partitioning and isolation mechanisms, enabling developers to select suitable configurations for different application requirements.

\noindent
\underline{\textbf{FreeRTOS.}} 
Fig.~\ref{agile_freertos} shows that 
Akira introduces only limited abstraction cost across different workloads, with an average slowdown of 1.10× compared with native FreeRTOS. 
More importantly, the overhead remains stable when additional isolation requirements are introduced, with MPU and privilege-switch configurations achieving 1.18× and 1.22× slowdown, respectively. 
The overhead mainly depends on workload execution granularity. Compute-intensive workloads such as \texttt{huffbench} and \texttt{matmult-int} reduce the relative impact of cross-domain communication and incur less than 5\% overhead, while workloads with shorter execution times such as \texttt{crc32} and \texttt{statemate} are more affected by domain transition latency.

Furthermore, we compare the modification effort required to support different component organizations and isolation requirements when directly modifying FreeRTOS and Akira (applied on FreeRTOS). 
As shown in Tab.~\ref{tab:freertos_agility_kloc}, Akira introduces a one-time framework implementation cost, including Fabric, Bridge, and Controller, which can be reused across subsequent system evolutions. 
After this initial cost, further customization, such as binding or switching isolation backends (e.g., MPU or privilege switching), is confined to Akira's specialization layer and specified through \texttt{config.yaml}, resulting in only 0.28 KLOC of modifications on average while leaving the underlying FreeRTOS implementation unchanged.
In contrast, achieving the same evolution directly in FreeRTOS requires 1.38 KLOC of intrusive modifications on average to kernel and subsystem source files. 
Specifically, as shown in Tab.~\ref{tab:freertos_modified_pct}, extending FreeRTOS with MPU support, privilege separation, I/O, or TCP functionality propagates changes across kernel components, subsystem implementations, and build scripts (e.g., Makefile), resulting in cascading modifications that affect up to 25\% of the codebase (Sec.~\ref{obs2}).

Overall, Akira enables agile system evolution by supporting workload-driven component reorganization and isolation configuration, allowing developers to explore different security–performance trade-offs without redesigning the underlying OS architecture.

\subsection{Performance Analysis} 
We design a set of micro-benchmarks~\cite{gu2020harmonizing} to evaluate the performance overhead introduced by Akira. 
In each benchmark, two isolated domains are configured (Akira realizes the domains using MPK-based isolation, while other systems treat their corresponding isolation boundaries as domains for comparison). 
We use \texttt{rdtsc} to measure the round-trip latency of an empty invocation, capturing the end-to-end cost of crossing the isolation boundary in each architecture.   


\begin{figure}[ht]
    \centering
    \captionsetup{skip=4pt}
    \begin{minipage}{0.49\linewidth}
        \centering
        \includegraphics[width=\linewidth]{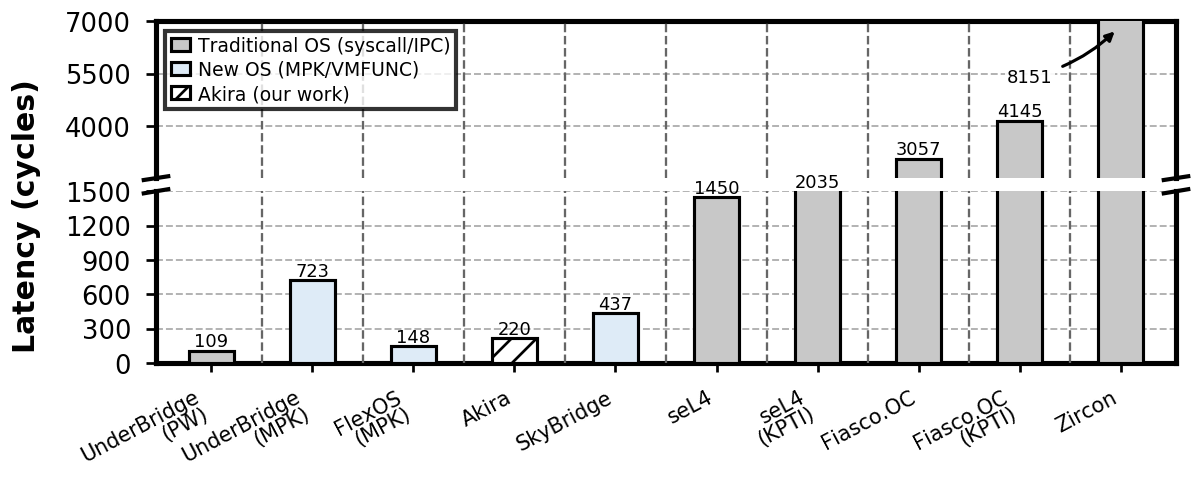}
        \vspace{-1.5em}
        \caption{Cross-domain round-trip latency. Gray: traditional OSs; blue: new OSs with hardware-assisted domain (MPK/VMFUNC); hatched: Akira (our work).}
        \label{cross-server}
    \end{minipage}
    \hfill
    \begin{minipage}{0.48\linewidth}
        \centering
        \includegraphics[width=\linewidth]{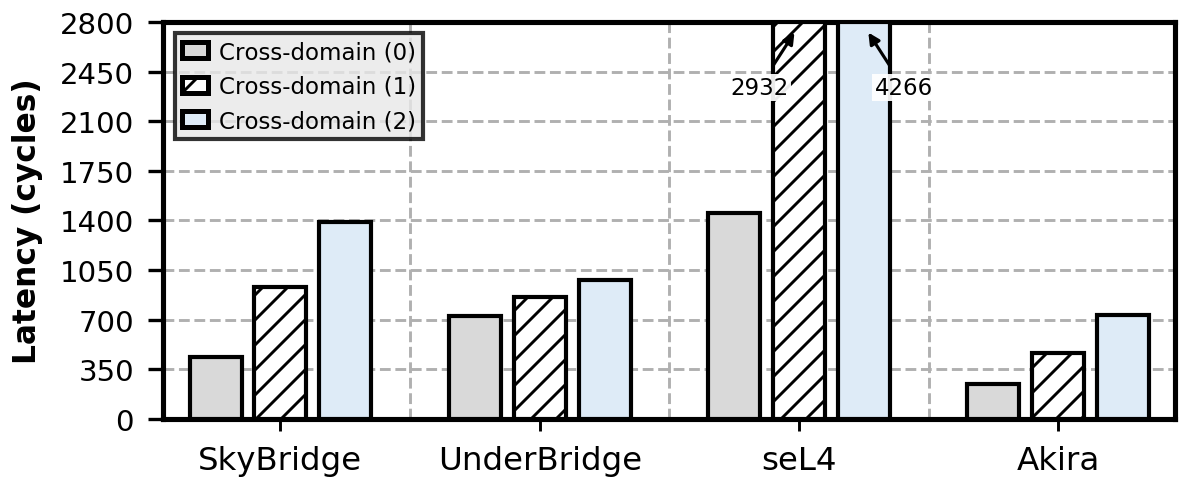}
        \caption{Latency of cross-domain round-trip as the call chain  increases. Gray, hatched, and light-blue denote 0, 1, and 2 intermediate domain hops. }
        \label{cross-domain}
    \end{minipage}
    \vspace{-5pt}
\end{figure} 

\underline{\textbf{Cross-domain IPC.} } 
Since Akira and microkernels both execute components in isolated protection domains and rely on explicit cross-domain communication, we use microkernels as baselines to evaluate Akira's communication efficiency. 
As shown in Fig.~\ref{cross-server}, for microkernels, services communicate through kernel-mediated IPC, where the latency is mainly determined by the complexity of the IPC path.  Akira's cross-domain round-trip is more than $6\times$ faster than seL4 and up to $37\times$ faster than Zircon. 
This advantage comes from Akira's configurable isolation and efficient communication path: BIU selects suitable isolation backends and, in this experiment, uses MPK-based bridge gates to switch protection states without privilege transitions. 
The Fabric path further avoids the heavyweight processing required by microkernel IPC.

We further compare Akira with OS designs that leverage hardware-assisted isolation, including SkyBridge, UnderBridge, and FlexOS. 
Compared to SkyBridge (EPT-based) and UnderBridge's user-to-kernel transition path (Privilege-based), Akira achieves a $2.0\times$--$3.3\times$ speedup by avoiding additional kernel-mediated IPC processing after domain switching. 
UnderBridge's kernel-to-kernel path (MPK-based) and FlexOS (MPK-based) achieve lower domain-switching overhead. FlexOS also supports multiple isolation mechanisms, including EPT and SGX. 
However, their isolation designs rely on existing component boundaries provided by the underlying OS architecture.  
Akira provides an OS architecture that enables systems with different structures to decompose, isolate, and compose components through a unified framework.

\underline{\textbf{Cross-domain Chain.} }
We further evaluate cross-domain invocations across multiple domains by varying the number of intermediate hops ($0$, $1$, and $2$).  
Fig.~\ref{cross-domain} compares the performance of seL4, SkyBridge (applied on Fiasco.OC), UnderBridge (applied on ChCore) and Akira (applied on Unikraft).  
Microkernel IPC accumulates substantial overhead as each hop repeats the IPC path; for example, seL4 latency increases by \(2.9\times\) with two intermediate hops. 
In contrast, hardware-assisted approaches and Akira introduce smaller per-hop overhead, enabling more scalable cross-domain communication.
Akira achieves the lowest latency across all configurations, outperforming SkyBridge by \(56\%\) with one intermediate hop and UnderBridge by \(54\%\) with two hops. 

Overall, Akira achieves low-latency cross-domain communication while preserving the flexibility to organize and interact with isolated components under different isolation configurations.

\subsection{Analysis of Performance Trade-offs} 
\begin{figure}[htpb]
    \centering
    \begin{minipage}{0.6\linewidth}
        \centering
        \includegraphics[width=0.9\linewidth]{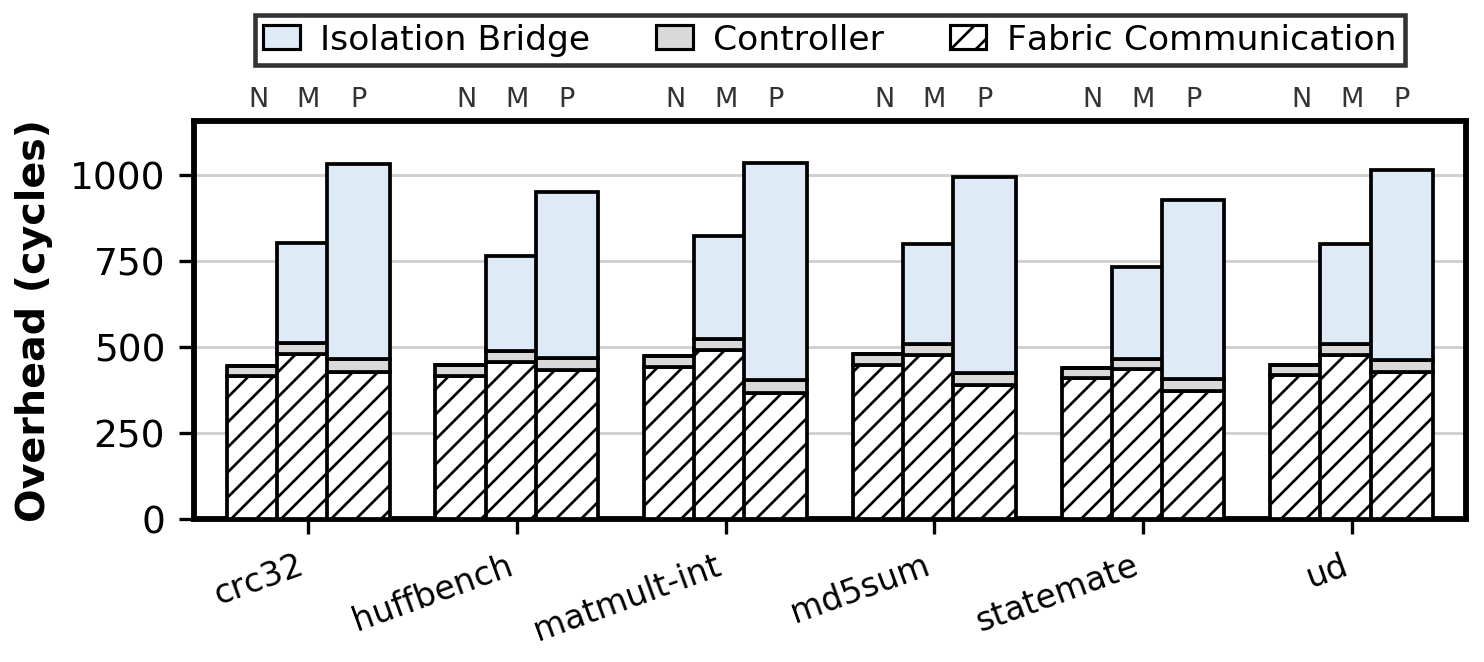}
        \vspace{-1.0em}
        \caption{Overhead of Fabric, Controller, and Bridge under None (\texttt{N}), MPU (\texttt{M}), and privilege-switch (\texttt{P}) backends
across Embench.}
        \label{trade-offs}
    \end{minipage}
    \hfill
    \begin{minipage}{0.35\linewidth}
        \centering
        \includegraphics[width=0.95\linewidth]{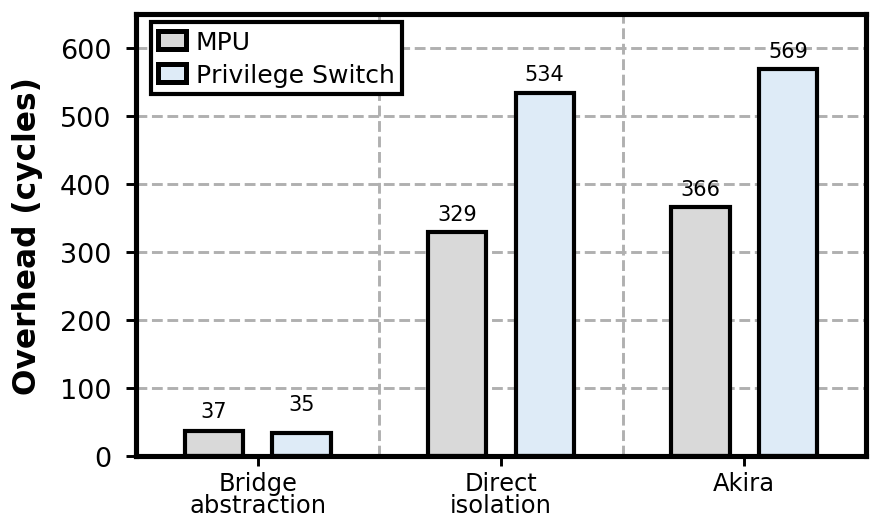}
        \caption{Breakdown of isolation transition overhead in Akira.}
        \label{ablation}
    \end{minipage}
\end{figure}

Fig.~\ref{trade-offs} quantifies the accumulated performance overhead across different execution-path stages in Akira (applied on FreeRTOS) running on Embench, revealing the architectural trade-off between isolation mechanisms and communication paths identified in Sec.~\ref{implementation}.

\begin{table}[htpb]
\centering
\vspace{-1.0em}
\caption{Two-hop invocation latency under different designs. 
$\mathrm{lib1}$, $\mathrm{lib2}$, and $\mathrm{lib3}$ are empty Unikraft components. $\mathrm{lib1}$ is placed in one isolation domain, while $\mathrm{lib2}$ and $\mathrm{lib3}$ share another.}
\vspace{-0.5em}
\label{tab:fabric_alab}
\footnotesize
\begin{tabular}{@{}llcrrr@{}}
\toprule
\textbf{Design} & \textbf{Hop} & \textbf{Boundary} & \textbf{Mechanism} & \textbf{Cycles} & \textbf{Total} \\
\midrule
\multirow{2}{*}{\shortstack[l]{All direct\\{\footnotesize(monolithic-style)}}}
  & lib1\,$\rightarrow$\,lib2 & Cross-domain & Privilege switch
    & 150 & \multirow{2}{*}{\textbf{175}} \\
  & lib2\,$\rightarrow$\,lib3 & Intra-domain & Direct call
    & 25 & \\
\midrule
\multirow{2}{*}{\shortstack[l]{Akira\\{\footnotesize(message $+$ direct)}}}
  & lib1\,$\rightarrow$\,lib2 & Cross-domain & Message $+$ privilege
    & 990 & \multirow{2}{*}{\textbf{1015}} \\
  & lib2\,$\rightarrow$\,lib3 & Intra-domain & Direct call
    & 25 & \\
\midrule
\multirow{2}{*}{\shortstack[l]{All message\\{\footnotesize(microkernel-style)}}}
  & lib1\,$\rightarrow$\,lib2 & Cross-domain & Message $+$ privilege
    & 990 & \multirow{2}{*}{\textbf{1980}} \\
  & lib2\,$\rightarrow$\,lib3 & Cross-domain & Message $+$ privilege
    & 990 & \\
\bottomrule
\end{tabular}
\end{table}

\underline{\textbf{Controller.} }
Across all configurations and workloads in Fig.~\ref{trade-offs}, Controller overhead stays low: about $6.7\%$--$6.8\%$ under None, $\sim$4.1\% under MPU, and $3.5\%$--$3.7\%$ under privilege switch.
As isolation strengthens, its relative share even shrinks, because domain-switch cost grows while routing, service-map lookup, and arbitration remain cheap.
Thus the control plane is not the bottleneck, confirming the lightweight controller design described in Sec.~\ref{implementation}.

\underline{\textbf{Bridge.}}
The Isolation Bridge overhead comprises two parts: (i)~the inherent cost of the underlying isolation mechanism and (ii)~the additional execution-path cost introduced by the AEU abstraction. 
Fig.~\ref{trade-offs} isolates (i) the overhead introduced by isolation mechanisms: changing only the isolation backend introduces about $36\%$ overhead under MPU and $51\%$--$61\%$ under privilege switching. 
Moreover, we analyze (ii): the additional execution-path cost introduced by the AEU abstraction mainly comes from the enter/leave operations during cross-domain communication.  
Fig.~\ref{ablation} isolates (ii): the bridge abstraction introduces only around 37 cycles of additional execution-path overhead, with negligible variation across isolation mechanisms. 
The result confirms that the abstraction layer adds a small and stable cost independent of the underlying isolation backend.

\underline{\textbf{Fabric.} }
Fig.~\ref{trade-offs} shows that Fabric communication dominates the overhead under None isolation, accounting for over $93\%$ across workloads. 
When MPU or privilege switching is enabled, isolation overhead reduces the contribution of Fabric communication, with its share decreasing to around $60\%$ and $40\%$, respectively. 
This indicates that, without isolation overhead, Fabric communication dominates; with stronger isolation, transition costs become the primary overhead.
To further analyze Fabric overhead, we evaluate a two-hop chain where only the first interaction crosses an isolation boundary (Tab.~\ref{tab:fabric_alab}).  
Routing all interactions by message mediation increases the cost by \(11.3\times\) compared with direct calls, while Akira incurs only \(5.8\times\) the cost by keeping intra-domain interactions as native calls. 
This shows that Akira confines Fabric overhead to necessary cross-domain interactions rather than applying message mediation to all component communication.  

Overall, the analysis shows that Akira introduces limited architectural overhead: the dominant costs stem from the underlying isolation mechanisms and unavoidable cross-domain transitions, while the controller, bridge abstraction, and optimized communication paths contribute only minor additional overhead.

\subsection{Scalability} 
We evaluate Akira scalability by varying Fabric layers, components per layer, and isolation domains to study the impact of topology size, component density, and isolation scale. 

\begin{figure}[htpb]
    \centering
    \begin{minipage}{0.49\linewidth}
        \centering
        \includegraphics[width=\linewidth]{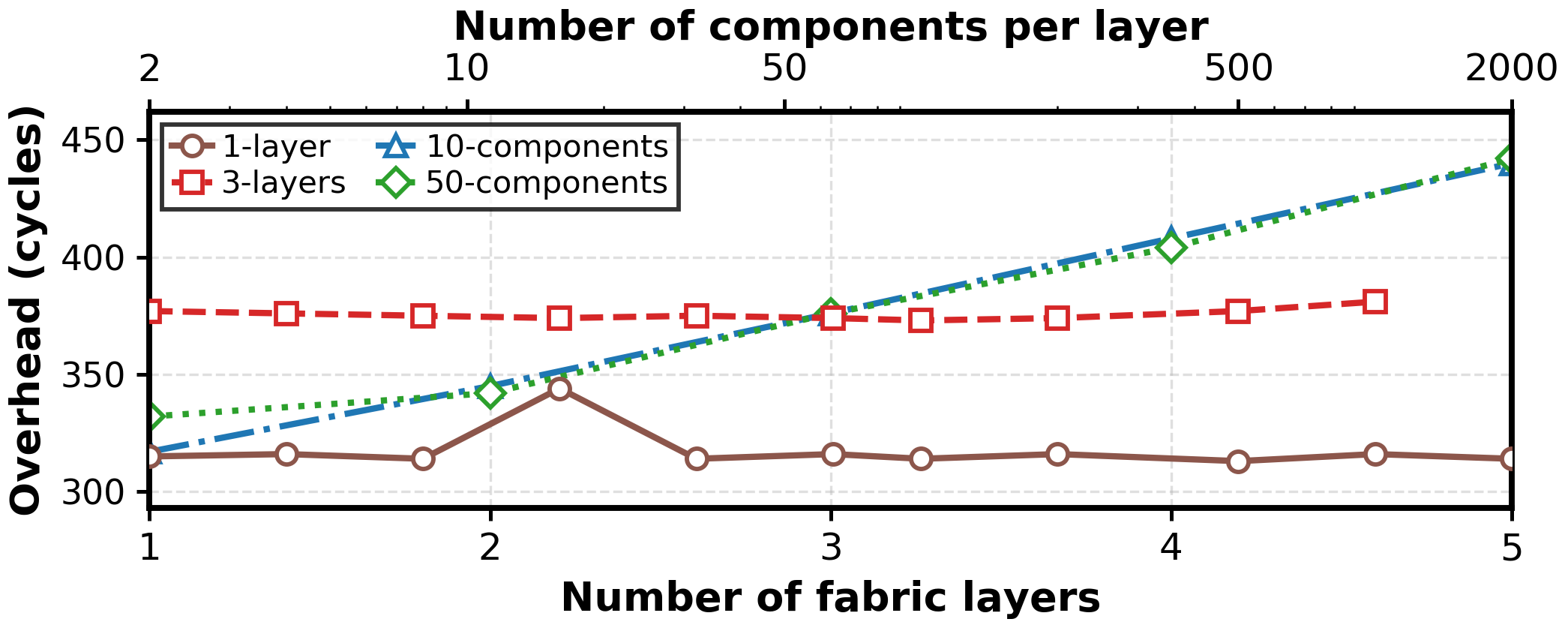}
        \vspace{-20pt}
        \caption{Fabric Scalability. Depth: scaling layers (1–5) with fixed layer width (10/50); width: scaling layer width (2–2,000) with fixed layers (1/3).} 
        \label{fabricsc}
    \end{minipage}\hfill
    \begin{minipage}{0.49\linewidth}
        \centering
        \includegraphics[width=0.96\linewidth]{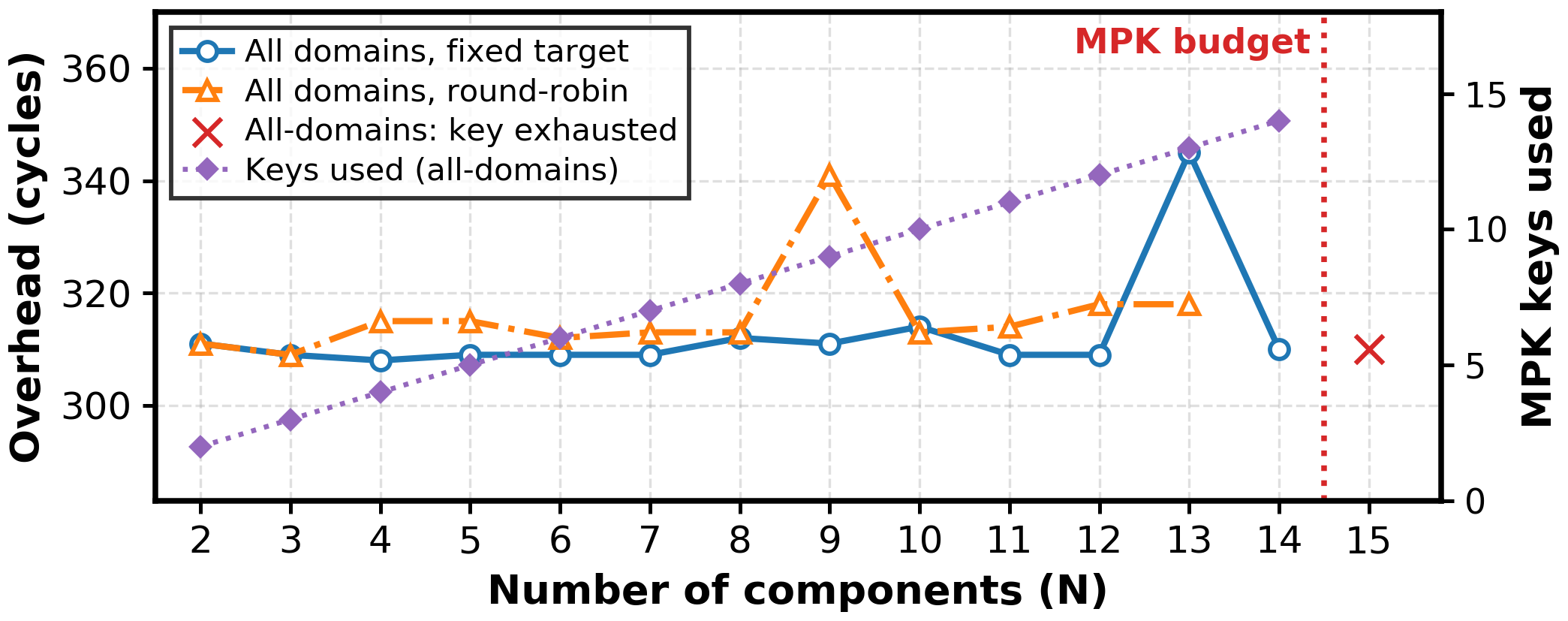}
        \vspace{-10pt}
        \caption{MPK Isolation-Domain Scalability. Fixed target: single-domain routing with a fixed key; round-robin: cyclic multi-domain routing with key rotation. } 
        \label{clusc}
    \end{minipage}
\end{figure}

\underline{\textbf{Fabric and Component.}} 
We evaluate Akira scalability by increasing the number of layers and components. 
Fig.~\ref{fabricsc} evaluates the impact of topology depth and width on Fabric overhead. We first increase the number of layers from 1 to 5 while holding 10 and 50 components per layer. The overhead grows approximately linearly with topology depth, from about 320~cycles at one layer to about 440~cycles at five layers, as each additional layer introduces another forwarding step in the communication path (about 30~cycles per additional layer). 
In contrast, increasing the number of components per layer from 2 to 2{,}000 has little impact on the cost: with a single layer the overhead remains around 310--320~cycles, and with three layers it stays near 375~cycles even at 2{,}000 components. This is because additional components only expand the available component set, while a \texttt{fabric\_gate} invocation follows the same forwarding path without extra lookup or processing. In the worst case (five layers), Fabric overhead remains below 450~cycles.
Therefore, Fabric scales well with topology width, while topology depth introduces a predictable linear cost.

\underline{\textbf{Isolation Domains.}} 
We choose Intel MPK as the isolation backend to evaluate how Akira scales with an increasing number of isolation domains under a hardware resource constraint. 
Intel MPK provides 16 protection keys, with two reserved for shared memory and the default domain, leaving 14 keys for component domains. 
To isolate the impact of domain count, we measure single-hop \texttt{fabric\_gate} invocations. 
Fig.~\ref{clusc} shows that cross-domain invocation latency remains stable as the number of domains increases, indicating that Akira's Fabric path does not introduce additional overhead with larger domain counts.  
This indicates that Akira's cross-domain invocation cost is insensitive to the number of domains, while scalability is primarily limited by the MPK key budget (extending to $N=15$ fails due to key exhaustion) rather than transition overhead. 
Thanks to the existing MPK key-extension techniques~\cite{mi2019skybridge,park2019libmpk}, Akira can multiplex protection domains onto a limited number of keys, enabling scalability beyond the native MPK domain limit.

Overall, these results indicate that Akira can be extended based on topology size and isolation requirements. 
Akira can independently manage component organization and isolate resources to support larger and more professional deployments. 


\section{Related Work}  
\label{sec:related}
\textbf{Modular and specialized OS architectures.} 
Prior work specializes OSs along two main axes (Sec.~\ref{section:motivation}): microkernels that enforce modularity via IPC and capabilities (e.g., seL4~\cite{klein2009sel4}, Barrelfish~\cite{baumann2009multikernel}, LegoOS~\cite{shan2018legoos}), and library OSs / unikernels that compose application-specific instances from reusable components (e.g., Exokernel~\cite{engler1995exokernel}, Unikraft~\cite{kuenzer2021unikraft}, MirageOS~\cite{madhavapeddy2013unikernels}, OSv~\cite{kivity2014osv}). 
Hybrid designs such as HongMeng~\cite{chen2024microkernel} and UnderBridge~\cite{gu2020harmonizing} combine reusable components with hardware-assisted isolation. 
Across these lines, however, component organization and isolation mechanisms remain largely predefined, limiting joint reconfiguration for evolving requirements. 

\textbf{Configurable Isolation for Component-based OSs.}
Prior work has explored isolation for component-based systems by placing components into protection domains and mediating cross-domain interactions by controlled transitions. 
Gate-based approaches rely on low-cost domain switching within a shared address space. 
For example, ERIM~\cite{vahldiek2019erim} and Hodor~\cite{hedayati2019hodor} use Intel MPK to isolate sensitive components, while CubicleOS~\cite{sartakov2021cubicleos} applies compartment-based isolation to library OS components. 
FlexOS~\cite{lefeuvre2022flexos} further explores configurable isolation by allowing different protection mechanisms (e.g., MPK, EPT, SGX) to be selected for components. 
SkyBridge~\cite{mi2019skybridge} and UnderBridge~\cite{gu2020harmonizing} optimize cross-domain transitions with hardware assistance, but remain coupled to the existing protection and communication paths. 
Classical process isolation and microkernel IPC~\cite{klein2009sel4} provide stronger domain separation, yet likewise bind interactions to predefined IPC interfaces. 
Tab.~\ref{tab:akira_vs_prior} summarizes the differences between Akira and other OSs in agile specialization. 
\begin{table}[ht]
\centering
\caption{Comparison with related systems.
\cmark:~supported;
$\circ$:~partial;
\xmark:~not supported.
\textit{Reusable}: components can be composed/reused without rewriting their internals.
\textit{Portable}: applicable to OSs with different component organizations.
\textit{Decoupled}: isolation policy can change without rewriting host component boundaries.}
\label{tab:akira_vs_prior}
\small
\setlength{\tabcolsep}{5pt}
\begin{tabular}{@{}lcccc@{}}
\toprule
 & \begin{tabular}[c]{@{}c@{}}Component\\reuse\end{tabular}
 & \begin{tabular}[c]{@{}c@{}}Portable\\across OSs\end{tabular}
 & \begin{tabular}[c]{@{}c@{}}Configurable\\isolation\end{tabular}
 & \begin{tabular}[c]{@{}c@{}}Isolation\\decoupled\end{tabular} \\
\midrule
CubicleOS~\cite{sartakov2021cubicleos}
 & \xmark & \xmark & \xmark & \xmark \\
UnderBridge~\cite{gu2020harmonizing}
 & \xmark & \xmark & $\circ$ & $\circ$ \\
HongMeng~\cite{chen2024microkernel}
 & $\circ$ & $\circ$ & $\circ$ & \xmark \\
FlexOS~\cite{lefeuvre2022flexos}
 & \xmark & \xmark & \cmark & $\circ$ \\
\midrule
\rowcolor{gray!12}
\textbf{Akira (Our work)}
 & \cmark & \cmark & \cmark & \cmark \\
\bottomrule
\end{tabular}
\end{table}

\textbf{Communication Abstraction and Configurable Interaction.}
Existing systems have explored communication abstraction to simplify interactions among distributed or modular components. 
At the OS level, systems such as Barrelfish~\cite{baumann2009multikernel} and LegoOS~\cite{shan2018legoos} adopt explicit communication mechanisms, such as message passing and remote service invocation, to improve communication modularity. 
However, their communication structures remain closely tied to predefined system organizations. 
Component middleware further abstracts communication relationships from component implementations. 
Frameworks such as AUTOSAR RTE~\cite{huang2022application}, OMG DDS~\cite{schlesselman2004omg}, CORBA~\cite{vinoski1997corba}, and ROS~\cite{quigley2009ros} use interface abstractions and runtime binding to manage component interactions independently of implementations. 
Similarly, HongMeng's distributed softbus provides unified abstractions for cross-device communication~\cite{chen2024microkernel}. 
However, these approaches mainly manage communication connectivity and routing, while execution domains and isolation mechanisms remain determined by the underlying system architecture. 
Software-defined networking (SDN) similarly separates programmable control logic from data forwarding~\cite{bannour2017distributed}. 
However, its control abstraction targets network traffic management rather than OS component communication and isolation.  
Therefore, existing approaches lack a unified mechanism for jointly adapting component interactions and isolation configurations in specialized OS environments.

\section{Conclusion}
We have presented Akira, a reconfigurable OS architecture that enables flexible component composition and isolation specialization across diverse system requirements. 
The design is guided by an in-depth analysis of current OS specialization bottlenecks, revealing that limited component abstraction and rigid isolation mechanisms hinder efficient adaptation to heterogeneous workloads. 
Based on these insights, Akira introduces a stable abstraction layer for components, separates system-level management from component functionality, and enables configurable isolation across diverse component organizations. 
Together, these mechanisms allow components and isolation policies to be composed and reconfigured without extensive modifications to existing implementations. 
We prototype Akira on FreeRTOS and Unikraft and demonstrate that it can rapidly adapt to diverse workload requirements while maintaining low communication and performance overhead.


\begin{acks}
We appreciate the anonymous reviewers for their helpful feedback. 
\end{acks}

\bibliographystyle{ACM-Reference-Format}
\bibliography{sample-base}

\end{document}